%% file: paper.tex
\documentclass[a4paper,11pt]{article}

\usepackage{hyperref}%
\hypersetup{%
  colorlinks=true,%
  linkcolor=blue,%
  citecolor=blue,%
  urlcolor=black,%
  bookmarksnumbered=true,%
  bookmarkstype=toc,%
  bookmarksopen=false,%
  pdftitle={Efimov effect for two particles on a semi-infinite line},%
  pdfkeywords={two-body problem; Efimov effect; discrete scale invariance; exact solutions},%
  pdfauthor={Satoshi Ohya}}%

\usepackage[margin=1in]{geometry}%

\usepackage[tt=false,sb]{libertine}%
\usepackage[T1]{fontenc}%
\usepackage{libertinust1math}%
\usepackage[cal=stix,scr=boondoxo,bb=boondox]{mathalpha}%
\usepackage{bm}%
\DeclareMathOperator{\e}{e}%
\allowdisplaybreaks[1]%

\usepackage{graphicx}%
\usepackage{epic,eepic,xcolor}%

\usepackage[font=small,hang,labelfont=bf]{caption}%

\usepackage[title]{appendix}

\usepackage[backend=biber,style=phys,biblabel=brackets,block=ragged,eprint=true,url=true]{biblatex}%
\title{\Large\bfseries Efimov effect for two particles on a semi-infinite line}%
\author{\normalsize Satoshi Ohya\\[1em]
  \small\itshape Institute of Quantum Science, Nihon University,\\
  \small\itshape Kanda-Surugadai 1-8-14, Chiyoda, Tokyo 101-8308, Japan\\[1ex]
  \small\ttfamily ohya.satoshi@nihon-u.ac.jp}%
\date{\small(Dated: \today)}%

\begin{document}
\maketitle%
\flushbottom%

\begin{abstract}
  The Efimov effect (in a broad sense) refers to the onset of a
  geometric sequence of many-body bound states as a consequence of the
  breakdown of continuous scale invariance to discrete scale
  invariance. While originally discovered in three-body problems in
  three dimensions, the Efimov effect has now been known to appear in
  a wide spectrum of many-body problems in various dimensions. Here we
  introduce a simple, exactly solvable toy model of two identical
  bosons in one dimension that exhibits the Efimov effect. We consider
  the situation where the bosons reside on a semi-infinite line and
  interact with each other through a pairwise $\delta$-function
  potential with a particular position-dependent coupling strength
  that makes the system scale invariant. We show that, for
  sufficiently attractive interaction, the bosons are bound together
  and a new energy scale emerges. This energy scale breaks continuous
  scale invariance to discrete scale invariance and leads to the onset
  of a geometric sequence of two-body bound states. We also study the
  two-body scattering off the boundary and derive the exact reflection
  amplitude that exhibits a log-periodicity. This article is intended
  for students and non-specialists interested in discrete scale
  invariance.
\end{abstract}

\newpage
\section{Introduction}
\label{section:1}
In his seminal paper in 1970, Efimov considered three identical bosons
with short-range pairwise interactions \cite{Efimov:1970zz}. He
pointed out that, when the two-body scattering length diverges, an
infinite number of three-body bound states appear, with energy levels
$\{E_{n}\}$ forming a geometric sequence. This phenomenon---generally
known as the Efimov effect---has attracted much attention because the
ratio $E_{n+1}/E_{n}\approx1/(22.7)^{2}$ is independent of the details
of the interactions as well as of the nature of the particles: it is
universal. More than thirty-five years after its prediction, this
effect was finally observed in cold atom experiments
\cite{Kraemer:2005,Zaccanti:2009,Gross:2009,Pollack:2009,Huang:2014},
which has triggered an explosion of research on the Efimov effect. For
more details, see the reviews
\cite{Nielsen:2001,Braaten:2004rn,Naidon:2016dpf,Greene:2017cik,DIncao:2017}.
(See also Refs.~\cite{Ferlaino:2010,Greene:2010,Bhaduri:2010} for a
more elementary exposition.)

Aside from its universal eigenvalues ratio, the Efimov effect takes
its place among the greatest theoretical discoveries in modern physics
because it was the first quantum many-body phenomenon to demonstrate
\textit{discrete scale invariance}---an invariance under enlargement
or reduction in the system size by a single scale factor
\cite{Sornette:1997pb}. It is now known that the emergence of a
geometric sequence in the bound states' discrete energies is
associated with the breakdown of continuous scale invariance to
discrete scale invariance \cite{Ovdat:2019ywg}, and can be found in a
wide spectrum of quantum many-body problems in various dimensions
\cite{Nishida:2009pg,Nishida:2011ew,Nishida:2012by,Moroz:2015,Ovdat:2017lho,Ohya:2021hju,Nishida:2021emz}.
The notion of the Efimov effect has therefore now been broadened to
include those generalizations, so that its precise meaning varies in
the literature. In the present paper, we will use the term ``Efimov
effect'' to simply refer to the onset of a geometric sequence in the
energies of many-body bound states as a consequence of the breakdown
of continuous scale invariance to discrete scale invariance.

To date, there exist several theoretical approaches to study the
Efimov effect. The most common approach is to directly analyze the
many-body Schr\"{o}dinger equation, which normally involves the use of
Jacobi coordinates, hyperspherical coordinates, the adiabatic
approximation, and the Faddeev equation \cite{Nielsen:2001}. Another
popular approach is to use second quantization, or quantum field
theory \cite{Braaten:2004rn}. Though the problem itself is
conceptually simple, it is hard for students and non-specialists to
master these techniques and to work out the physics of the Efimov
effect. The essential part of this phenomenon, however, can be
understood from undergraduate-level quantum mechanics without using
any fancy techniques.

\begin{figure}[t]
  \centering%
  \input{figure1.eepic}%
  \caption{Position dependence of the coupling strength $g(x)$.}
  \label{figure:1}
\end{figure}
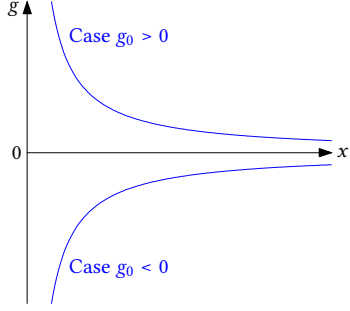

This paper is aimed at introducing a simple toy model for a two-body
system that exhibits the Efimov effect. We consider two identical
bosons on the half-line $\mathbb{R}_{+}=\{x:x\geq0\}$ with a pairwise
$\delta$-function interaction. The Hamiltonian of such a system is
given by
\begin{align}
  H=-\frac{\hbar^{2}}{2m}\left(\frac{\partial^{2}}{\partial x_{1}^{2}}+\frac{\partial^{2}}{\partial x_{2}^{2}}\right)+g(x_{1})\delta(x_{1}-x_{2}),\label{eq:1}
\end{align}
where $m$ is the mass of each particle and $x_{j}\in\mathbb{R}_{+}$
($j=1,2$) is the coordinate of the $j$th particle. Here $g(x)$ is a
coupling strength. In this paper, we will focus on the
position-dependent coupling strength that satisfies the scaling law
$g(\e^{t}x)=\e^{-t}g(x)$, where $t$ is an arbitrary real
number. Notice that, up to an overall constant factor, this scaling
law has a unique solution $g(x)\propto1/x$. For the following
discussion, it is convenient to choose
\begin{align}
  g(x)=\frac{\hbar^{2}}{m}\frac{g_{0}}{x},\label{eq:2}
\end{align}
where $g_{0}$ is a dimensionless real number that can either be
positive or negative. Physically, Eq.~\eqref{eq:2} models the
situation where the interaction strength becomes stronger as the
particles come closer to the boundary $x_{1}=x_{2}=0$ (see
Fig.~\ref{figure:1}). This two-body interaction is essentially
equivalent to the so-called \textit{scaling trap} introduced in
Ref.~\cite{Nishida:2012by}, where the Efimov effect was discussed in
the context of two non-identical particles on the whole line
$\mathbb{R}$. As we will see shortly, our two-identical-particle
problem on $\mathbb{R}_{+}$ enjoys simple solutions and is more
tractable than the corresponding two-non-identical-particle problem on
$\mathbb{R}$.

\pagebreak[4]

The rest of the paper is devoted to the detailed analysis of the
eigenvalue problem of $H$. Before going into details, however, it is
worth summarizing the symmetry properties of the model. Of particular
importance are the following:
\begin{itemize}
\item\textit{Permutation invariance.} Thanks to the relation
  $g(x_{1})\delta(x_{1}-x_{2})=g(x_{2})\delta(x_{2}-x_{1})$, the
  Hamiltonian (1) is invariant under the permutation of coordinates,
  $(x_{1},x_{2})\mapsto(x_{2},x_{1})$. Note that this permutation
  invariance is necessary for Eq.~\eqref{eq:1} to be a Hamiltonian of
  indistinguishable particles, where, for bosons, the two-body
  wavefunction should satisfy $\psi(x_{1},x_{2})=\psi(x_{2},x_{1})$.
  We will see in Sec.~\ref{section:3.1} that this invariance greatly
  simplifies the analysis.
\item\textit{Scale invariance.} Thanks to the relations
  $g(\e^{t}x_{1})=\e^{-t}g(x_{1})$ and
  $\delta(\e^{t}x_{1}-\e^{t}x_{2})=\e^{-t}\delta(x_{1}-x_{2})$, the
  Hamiltonian \eqref{eq:1} transforms as $H\mapsto \e^{-2t}H$ under
  the scale transformation
  $(x_{1},x_{2})\mapsto(\e^{t}x_{1},\e^{t}x_{2})$. This transformation
  law has significant implications for the spectrum of $H$. Let
  $\psi_{E}(x_{1},x_{2})$ be a solution to the eigenvalue equation
  $H\psi_{E}(x_{1},x_{2})=E\psi_{E}(x_{1},x_{2})$. Then,
  $\psi_{E}(\e^{t}x_{1},\e^{t}x_{2})$ automatically satisfies
  $H\psi_{E}(\e^{t}x_{1},\e^{t}
  x_{2})=\e^{2t}E\psi_{E}(\e^{t}x_{1},\e^{t}x_{2})$; that is,
  $\psi_{E}(\e^{t}x_{1},\e^{t}x_{2})$ is proportional to the
  eigenfunction $\psi_{\e^{2t}E}(x_{1},x_{2})$ corresponding to the
  eigenvalue $\e^{2t}E$. The proportionality coefficient can be
  determined by requiring that both $\psi_{E}$ and $\psi_{\e^{2t}E}$
  be normalized. The result is the following scaling law:
  \begin{equation}
    \psi_{\e^{2t}E}(x_{1},x_{2})=\e^{t}\psi_{E}(\e^{t}x_{1},\e^{t}x_{2}).\label{eq:3}
  \end{equation}
  If this indeed holds for any $t\in\mathbb{R}$, $\e^{2t}E$ can take
  any arbitrary (positive) value so that the spectrum of $H$ is
  continuous. As we will see in Sec.~\ref{section:3.2}, however, if
  $g_{0}$ is smaller than a critical value $g_{\ast}$,
  Eq.~\eqref{eq:3} holds only for some discrete
  $t\in t_{\ast}\mathbb{Z}=\{0,\pm t_{\ast},\pm2t_{\ast},\cdots\}$;
  that is, continuous scale invariance is broken to discrete scale
  invariance, defined by a characteristic scale $t_{\ast}$. As a
  consequence, there appears a geometric sequence of (negative) energy
  eigenvalues,
  $\{E_{0},E_{0}\e^{\pm 2t_{\ast}},E_{0}\e^{\pm4t_{\ast}},\cdots\}$,
  where $E_{0}(<0)$ is a newly emergent energy scale. One of the goal
  of this paper is to show this result using only undergraduate-level
  calculus.
\end{itemize}

It should be noted that there is no translation invariance in our
model: it is explicitly broken by the boundary at $x=0$ as well as by
the position-dependent coupling strength \eqref{eq:2}. This
non-invariance means that the total momentum---the canonical conjugate
of the center-of-mass coordinate---is not a well-defined conserved
quantity. In other words, the two-body wavefunction cannot be of the
separation-of-variable form $\psi(x_{1},x_{2})=\e^{iPX/\hbar}\phi(x)$,
where $X=(x_{1}+x_{2})/2$ is the center-of-mass coordinate, $P$ the
total momentum, $x=x_{1}-x_{2}$ the relative coordinate, and $\phi$
the wavefunction of relative motion. In the next section, we will
first introduce an alternative coordinate system that is more suitable
for the two-body problem on the half-line $\mathbb{R}_{+}$, before
solving the problem in Sec.~\ref{section:3}.

\section{Two-body problem without translation invariance}
\label{section:2}
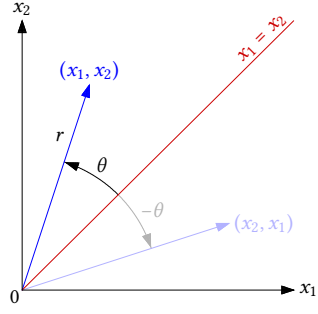
\begin{figure}[t]
  \centering%
  \input{figure2.eepic}%
  \caption{Polar coordinates $(r,\theta)$ in the
    $(x_{1},x_{2})$-space. The red line represents the set of two-body
    coincidence points.}
  \label{figure:2}
\end{figure}

Let us first introduce a new coordinate system in the
$(x_{1},x_{2})$-space. In what follows, we will work with the polar
coordinate system $(r,\theta)$ defined as follows (see
Fig.~\ref{figure:2}):
\begin{subequations}
  \begin{align}
    x_{1}&=r\cos(\theta+\tfrac{\pi}{4}),\label{eq:4a}\\
    x_{2}&=r\sin(\theta+\tfrac{\pi}{4}),\label{eq:4b}
  \end{align}
\end{subequations}
or, equivalently,
\begin{subequations}
  \begin{align}
    r&=\sqrt{x_{1}^{2}+x_{2}^{2}},\label{eq:5a}\\
    \theta&=\frac{1}{2i}\log\left(\frac{x_{1}+ix_{2}}{x_{1}-ix_{2}}\right)-\frac{\pi}{4},\label{eq:5b}
  \end{align}
\end{subequations}
where $r\in[0,\infty)$ and $\theta\in[-\pi/4,\pi/4]$. Note that
$\theta=0$ and $\theta=\pm\pi/4$ correspond to the two-body
coincidence point $x_{1}=x_{2}$ and to the boundaries $x_{1}=0$ and
$x_{2}=0$, respectively. Note also that the permutation
$(x_{1},x_{2})\mapsto(x_{2},x_{1})$ corresponds to the parity
transformation $(r,\theta)\mapsto(r,-\theta)$; see
Fig.~\ref{figure:2}.

In the coordinate system $(r,\theta)$, the kinetic energy part of the
two-body Hamiltonian \eqref{eq:1} takes the following form:
\begin{align}
  H_{0}
  &=-\frac{\hbar^{2}}{2m}\left(\frac{\partial^{2}}{\partial x_{1}^{2}}+\frac{\partial^{2}}{\partial x_{2}^{2}}\right)\nonumber\\
  &=-\frac{\hbar^{2}}{2m}\left(\frac{1}{r}\frac{\partial}{\partial r}r\frac{\partial}{\partial r}+\frac{1}{r^{2}}\frac{\partial^{2}}{\partial\theta^{2}}\right)\nonumber\\
  &=\frac{\hbar^{2}}{2m}r^{-\frac{1}{2}}\left(-\frac{\partial^{2}}{\partial r^{2}}+\frac{-\partial_{\theta}^{2}-\frac{1}{4}}{r^{2}}\right)r^{\frac{1}{2}},\label{eq:6}
\end{align}
where $\partial_{\theta}=\partial/\partial\theta$. Likewise, the
potential energy part is rewritten as
\begin{align}
  V(x_{1},x_{2})
  &=\frac{\hbar^{2}}{m}\frac{g_{0}}{x_{1}}\delta(x_{1}-x_{2})\nonumber\\
  &=\frac{\hbar^{2}}{m}\frac{g_{0}}{r\cos(\theta+\frac{\pi}{4})}\delta(r\cos(\theta+\tfrac{\pi}{4})-r\sin(\theta+\tfrac{\pi}{4}))\nonumber\\
  &=\frac{\hbar^{2}}{m}\frac{g_{0}}{\sqrt{2}r^{2}\cos(\frac{\pi}{4})}\delta(\theta)\nonumber\\
  &=\frac{\hbar^{2}}{m}\frac{g_{0}}{r^{2}}\delta(\theta)\quad\text{for $\theta\in[-\tfrac{\pi}{4},\tfrac{\pi}{4}]$}.\label{eq:7}
\end{align}
In the third equality we have used the formula
$\delta(f(\theta))=(1/|f^{\prime}(\theta_{0})|)\delta(\theta-\theta_{0})$,
where $f(\theta)=r\cos(\theta+\pi/4)-r\sin(\theta+\pi/4)$,
$f^{\prime}=df/d\theta$, and $\theta_{0}$ is a root of the equation
$f(\theta)=0$ for $\theta\in[-\pi/4,\pi/4]$ and given by
$\theta_{0}=0$. The total Hamiltonian $H=H_{0}+V$ can then be cast
into the following form:
\begin{equation}
  H=\frac{\hbar^{2}}{2m}r^{-\frac{1}{2}}\left(-\frac{\partial^{2}}{\partial r^{2}}+\frac{\Delta_{\theta}-\frac{1}{4}}{r^{2}}\right)r^{\frac{1}{2}},\label{eq:8}
\end{equation}
where
\begin{equation}
  \Delta_{\theta}=-\frac{\partial^{2}}{\partial\theta^{2}}+2g_{0}\delta(\theta).\label{eq:9}
\end{equation}

Now we are ready to analyze the Schr\"{o}dinger equation by means of
the separation of variables. Suppose that the two-body wavefunction is
of the form
\begin{equation}
  \psi(x_{1},x_{2})=r^{-\frac{1}{2}}R(r)\Theta(\theta).\label{eq:10}
\end{equation}
Let $\lambda$ be an eigenvalue of the operator $\Delta_{\theta}$.
Then, the time-independent Schr\"{o}dinger equation $H\psi=E\psi$ can
be reduced to the following set of differential equations:
\begin{subequations}
  \begin{align}
    \left(-\frac{d^{2}}{d\theta^{2}}+2g_{0}\delta(\theta)\right)\Theta(\theta)&=\lambda\Theta(\theta),\label{eq:11a}\\
    \left(-\frac{d^{2}}{dr^{2}}+\frac{\lambda-\frac{1}{4}}{r^{2}}\right)R(r)&=\frac{2mE}{\hbar^{2}}R(r).\label{eq:11b}
  \end{align}
\end{subequations}
The energy eigenvalues are determined by the inverse-square potential,
which has been widely studied over the years in the context of quantum
anomaly (symmetry breaking by quantization) or renormalization
\cite{Camblong:2000ec,Camblong:2000ax,Coon:2002sua,Bawin:2003dm,Braaten:2004pg,Hammer:2005sa,Essin:2006,Kaplan:2009kr},
and is known to support a geometric sequence of bound states if
$\lambda<0$ \cite{Case:1950an}. As we will see shortly, if $g_{0}$ is
smaller than a critical value $g_{\ast}$, the lowest eigenvalue
$\lambda_{0}$ in the eigenvalue equation \eqref{eq:11a} becomes
negative. Hence, in such a $\lambda_{0}$-channel, continuous scale
invariance can be broken down to discrete scale invariance, and the
two-body bound-state energies follow a geometric sequence. Let us next
see this by solving the differential equations \eqref{eq:11a} and
\eqref{eq:11b} explicitly.

\section{Two-body Efimov effect with boundary}
\label{section:3}
\subsection{Solution to the angular equation}
\label{section:3.1}
Let us first solve the angular equation \eqref{eq:11a}. To this end,
we need to specify the connection conditions at $\theta=0$ and the
boundary conditions at $\theta=\pm\pi/4$. We start with the connection
conditions equivalent to the $\delta$-function potential.

As is well known, the $\delta$-function potential system
\eqref{eq:11a} is equivalent to the differential equation
$-\Theta^{\prime\prime}=\lambda\Theta$ for $\theta\neq0$ with the
following connection conditions at $\theta=0$ \cite{Capri:1985}:
\begin{subequations}
  \begin{align}
    -\Theta^{\prime}(0_{+})+\Theta^{\prime}(0_{-})+g_{0}(\Theta(0_{+})+\Theta(0_{-}))=0,\label{eq:12a}\\
    \Theta(0_{+})=\Theta(0_{-}),\label{eq:12b}
  \end{align}
\end{subequations}
where the prime (${}^{\prime}$) indicates the derivative with respect
to $\theta$.

Let us next take into account the symmetry of the two-body
wavefunction. Since we are dealing with identical bosons, the
wavefunction must be symmetric under the permutation
$\psi(x_{1},x_{2})=\psi(x_{2},x_{1})$.\footnote{For fermions, we have
  $\psi(x_{1},x_{2})=-\psi(x_{2},x_{1})$, which is equivalent to
  $\Theta(\theta)=-\Theta(-\theta)$. The connection conditions
  \eqref{eq:12a} and \eqref{eq:12b} are then reduced to the Dirichlet
  boundary conditions $\Theta(0_{\pm})=0$, in which case $\lambda$
  cannot be negative (if we impose $\Theta(\pm\pi/4)=0$). Hence, for
  fermions, the Efimov effect cannot be realized with the Hamiltonian
  \eqref{eq:1}. In order to realize the Efimov effect for fermions
  with a pairwise contact interaction, one has to use the Hamiltonian
  $H=-(\hbar^{2}/(2m))(\partial^{2}/\partial
  x_{1}^{2}+\partial^{2}/\partial
  x_{2}^{2})+(\hbar^{2}/m)\varepsilon(x_{1}-x_{2};x_{1}/g_{0})$, where
  $\varepsilon$ is the so-called $\varepsilon$-function potential
  defined by
  $\varepsilon(x;c)=\lim_{a\to0_{+}}(1/(2c)-1/a)(\delta(x+a)+\delta(x-a))$.
  For simplicity, in this paper we will not touch upon the fermionic
  case. For more details of the $\varepsilon$-function potential, see
  Refs.~\cite{Cheon:1997rx,Cheon:1998iy}.}  In the polar coordinate
system, this is equivalent to $\Theta(\theta)=\Theta(-\theta)$, whose
derivative gives $\Theta^{\prime}(\theta)=-\Theta^{\prime}(-\theta)$.
Hence, at the two-body coincidence point $\theta=0$, there must hold
the following additional conditions:
\begin{equation}
  \Theta(0_{+})=\Theta(0_{-})
  \quad\text{and}\quad
  \Theta^{\prime}(0_{+})=-\Theta^{\prime}(0_{-}).\label{eq:13}
\end{equation}
Thus, for identical bosons, Eq.~\eqref{eq:12a} can be reduced to the
following Robin boundary conditions \cite{Gustafson:1998}:
\begin{equation}
  \mp\Theta^{\prime}(0_{\pm})+g_{0}\Theta(0_{\pm})=0.\label{eq:14}
\end{equation}

Let us now specify the boundary conditions at $\theta=\pm\pi/4$. For
simplicity, we will impose the following Dirichlet boundary
conditions:\footnote{Alternatively, one can impose, e.g., the Neumann
  boundary conditions $\Theta^{\prime}(\pm\pi/4)=0$, in which case the
  critical value is $g_{\ast}=0$.}
\begin{equation}
  \Theta(\pm\tfrac{\pi}{4})=0.\label{eq:15}
\end{equation}

\begin{figure}[t]
  \centering%
  \input{figure3.eepic}%
  \caption{Left: $g_{0}$ dependence of
    $\{\lambda_{0},\lambda_{1},\lambda_{2},\cdots\}$. The lowest
    eigenvalue $\lambda_{0}$ becomes negative if $g_{0}$ goes below
    the critical value $g_{\ast}=-4/\pi$. Right: A typical profile of
    $\Theta_{\lambda_{0}}(\theta)$ for $g_{0}<g_{\ast}$.}
  \label{figure:3}
\end{figure}
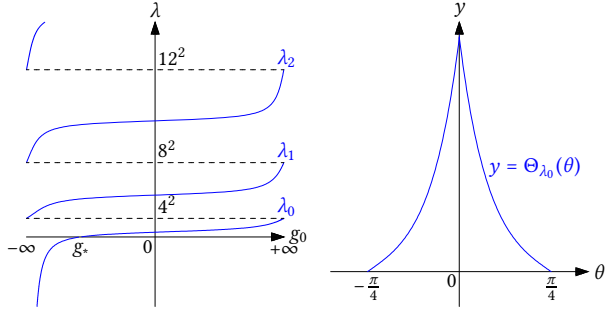

Now it is straightforward to solve the angular equation
\eqref{eq:11a}. Thanks to the property
$\Theta(\theta)=\Theta(-\theta)$, it is sufficient to solve the
differential equation $-\Theta^{\prime\prime}=\lambda\Theta$ in the
$0\leq\theta\leq\pi/4$ region under the boundary conditions
$-\Theta^{\prime}(0_{+})+g_{0}\Theta(0_{+})=0$ and
$\Theta(\pi/4)=0$. The resulting solution for $\lambda\neq0$ is
\begin{equation}
  \Theta_{\lambda}(\theta)=A_{\lambda}\sin\left(\sqrt{\lambda}\left(\frac{\pi}{4}-|\theta|\right)\right),\label{eq:16}
\end{equation}
where $A_{\lambda}$ is a normalization constant. Here $\lambda$ is a
root of the transcendental equation
\begin{equation}
  g_{0}=-\sqrt{\lambda}\cot\left(\frac{\pi}{4}\sqrt{\lambda}\right).\label{eq:17}
\end{equation}
For $\lambda<0$, the square root should be understood as
$\sqrt{\lambda}=i\sqrt{|\lambda|}$. In this case, the angular
wavefunction
$\Theta_{\lambda}(\theta)\propto\sinh(\sqrt{|\lambda|}(\pi/4-|\theta|))$
sharply localizes to the two-body coincidence point $\theta=0$; that
is, it describes a two-body bound state (see the right panel of
Fig.~\ref{figure:3}).

Though the transcendental equation \eqref{eq:17} cannot be solved
analytically, the $g_{0}$ dependence of its solutions can be seen by
plotting the graph of
$g_{0}=-\sqrt{\lambda}\cot(\pi\sqrt{\lambda}/4)$. As can be observed
in the left panel of Fig.~\ref{figure:3}, the lowest eigenvalue
$\lambda_{0}$ becomes negative for $g_{0}<g_{\ast}$, where $g_{\ast}$
is the critical value given by
\begin{equation}
  g_{\ast}:=\lim_{\lambda\to0}\left[-\sqrt{\lambda}\cot\left(\frac{\pi}{4}\sqrt{\lambda}\right)\right]=-\frac{4}{\pi}.\label{eq:18}
\end{equation}
Hence, in the $\lambda_{0}$-channel, continuous scale invariance must
be broken down to discrete scale invariance for $g_{0}<g_{\ast}$. Let
us next see this by solving the radial equation \eqref{eq:11b}.

\subsection{Boundary-localized two-body Efimov states}
\label{section:3.2}
From here on, we will consider the situation where $g_{0}<g_{\ast}$,
and focus on the case $E<0$ in the channel $\lambda=\lambda_{0}<0$. In
this case, there exists a square-integrable solution to the
differential equation \eqref{eq:11b} whose asymptotic behavior as
$r\to\infty$ is
$R_{\kappa\lambda}(r)\to N_{\kappa\lambda}\e^{-\kappa r}$, where
$N_{\kappa\lambda}$ is a normalization constant and
$\kappa=\sqrt{2m|E|/\hbar^{2}}>0$. The full solution is given
by\footnote{For $E=-\hbar^{2}\kappa^{2}/(2m)$ and $\lambda=-\nu^{2}$,
  Eq.~\eqref{eq:11b} is equivalent to the modified Bessel differential
  equation
  $(r^{2}d^{2}/dr^{2}+rd/dr-\kappa^{2}r^{2}+\nu^{2})(r^{-\frac{1}{2}}R)=0$. The
  two independent solutions to this equation are the modified Bessel
  functions of the first and second kind, $I_{i\nu}(\kappa r)$ and
  $K_{i\nu}(\kappa r)$, respectively, where the former is non-square
  integrable while the latter is square integrable. For more details
  of the modified Bessel functions, see, e.g.,
  Ref.~\cite{Watson:1995}.}
\begin{equation}
  R_{\kappa\lambda}(r)=N_{\kappa\lambda}\sqrt{\frac{2\kappa r}{\pi}}K_{i\nu}(\kappa r),\quad\nu=\sqrt{|\lambda|},\label{eq:19}
\end{equation}
where $K_{i\nu}$ is the modified Bessel function of the second kind
whose asymptotic behavior is
$K_{i\nu}(\kappa r)\to\sqrt{\pi/(2\kappa r)}\e^{-\kappa r}$ as
$\kappa r\to\infty$. Note that Eq.~\eqref{eq:19} together with
Eq.~\eqref{eq:16} describes a two-body wavefunction that localizes to
$\theta=0$ and $r=0$; that is, it describes the two-body bound state
that is localized to the boundary ($x_{1}=x_{2}=0$).

It should be noted that at this stage $\kappa$ is an arbitrary
positive real constant. To determine its possible values, we follow
the argument in Ref.~\cite{Case:1950an} and require the orthonormality
of the radial wavefunctions. Let $R_{\kappa\lambda}$ and
$R_{\kappa^{\prime}\lambda}$ be two distinct solutions to
Eq.~\eqref{eq:11b}. Then we have
\begin{subequations}
  \begin{align}
    -R_{\kappa\lambda}^{\prime\prime}+\frac{\lambda-\frac{1}{4}}{r^{2}}R_{\kappa\lambda}&=-\kappa^{2}R_{\kappa\lambda},\label{eq:20a}\\
    -\overline{R_{\kappa^{\prime}\lambda}^{\prime\prime}}+\frac{\lambda-\frac{1}{4}}{r^{2}}\overline{R_{\kappa^{\prime}\lambda}}&=-\kappa^{\prime2}\overline{R_{\kappa^{\prime}\lambda}},\label{eq:20b}
  \end{align}
\end{subequations}
where the overline ($\overline{\phantom{n}}$) stands for the complex
conjugate and the prime here indicates the derivative with respect to
$r$. By multiplying $\overline{R_{\kappa^{\prime}\lambda}}$ to
Eq.~\eqref{eq:20a} and $R_{\kappa\lambda}$ to Eq.~\eqref{eq:20b} and
then subtracting one from the other, we get
\begin{align}
  (-\kappa^{\prime2}+\kappa^{2})\overline{R_{\kappa^{\prime}\lambda}}R_{\kappa\lambda}
  &=\overline{R_{\kappa^{\prime}\lambda}}R_{\kappa\lambda}^{\prime\prime}-\overline{R_{\kappa^{\prime}\lambda}^{\prime\prime}}R_{\kappa\lambda}\nonumber\\
  &=\frac{d}{dr}\left(\overline{R_{\kappa^{\prime}\lambda}}R_{\kappa\lambda}^{\prime}-\overline{R_{\kappa^{\prime}\lambda}^{\prime}}R_{\kappa\lambda}\right).\label{eq:21}
\end{align}
By integrating both sides from $r=0$ to $\infty$, we find
\begin{align}
  (-\kappa^{\prime2}+\kappa^{2})\int_{0}^{\infty}\!\!\!dr\,\overline{R_{\kappa^{\prime}\lambda}(r)}R_{\kappa\lambda}(r)
  &=\int_{0}^{\infty}\!\!\!dr\,\frac{d}{dr}\left(\overline{R_{\kappa^{\prime}\lambda}(r)}R_{\kappa\lambda}^{\prime}(r)-\overline{R_{\kappa^{\prime}\lambda}^{\prime}(r)}R_{\kappa\lambda}(r)\right)\nonumber\\
  &=-\lim_{r\to0}\left(\overline{R_{\kappa^{\prime}\lambda}(r)}R_{\kappa\lambda}^{\prime}(r)-\overline{R_{\kappa^{\prime}\lambda}^{\prime}(r)}R_{\kappa\lambda}(r)\right)\nonumber\\
  &=\frac{2\sqrt{\kappa\kappa^{\prime}}\sin(\nu\log\frac{\kappa}{\kappa^{\prime}})}{\sinh(\nu\pi)}\overline{N_{\kappa^{\prime}\lambda}}N_{\kappa\lambda},\label{eq:22}
\end{align}
where the second equality follows from
$R_{\kappa\lambda},R_{\kappa^{\prime}\lambda}\to0$ in the limit
$r\to\infty$, and the last equality follows from the short-distance
behavior of the modified Bessel function (see Eq.~\eqref{eq:A4} in
Appendix \ref{appendix:A}).

\begin{figure}[t]
  \centering%
  \input{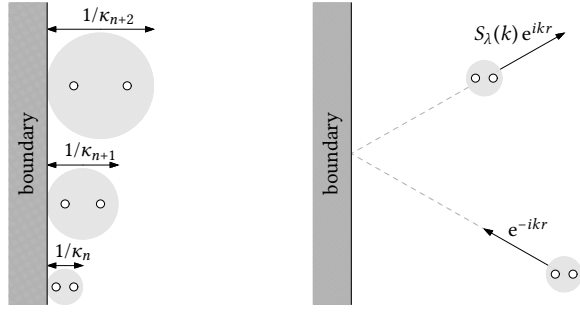}%
  \caption{Schematic pictures of the boundary-localized two-body bound
    states with different binding energies (left) and the scattering
    of a two-body bound state off the boundary (right). The boundary
    plays the role of the (infinitely heavy) third particle (see
    Appendix \ref{appendix:B} for details).}
  \label{figure:4}
\end{figure}

Now, Eq.~\eqref{eq:22} enables us to determine the normalization
constant as well as the energy eigenvalues. First, the normalization
constant is determined by requiring that $R_{\kappa\lambda}$ have the
unit norm:
\begin{align}
  1
  &=\int_{0}^{\infty}\!\!\!dr\,|R_{\kappa\lambda}(r)|^{2}\nonumber\\
  &=\lim_{\kappa^{\prime}\to\kappa}\int_{0}^{\infty}\!\!\!dr\,\overline{R_{\kappa^{\prime}\lambda}(r)}R_{\kappa\lambda}(r)\nonumber\\
  &=\lim_{\kappa^{\prime}\to\kappa}\frac{2\sqrt{\kappa\kappa^{\prime}}\sin(\nu\log\frac{\kappa}{\kappa^{\prime}})}{\sinh(\nu\pi)(\kappa^{2}-\kappa^{\prime2})}\overline{N_{\kappa^{\prime}\lambda}}N_{\kappa\lambda}\nonumber\\
  &=\frac{\nu}{\kappa\sinh(\nu\pi)}|N_{\kappa\lambda}|^{2},\label{eq:23}
\end{align}
where the last equality follows from
$\sin(\nu\log(\kappa/\kappa^{\prime}))=\sin(\nu\log(1+(\kappa-\kappa^{\prime})/\kappa^{\prime}))=\nu(\kappa-\kappa^{\prime})/\kappa^{\prime}+O((\kappa-\kappa^{\prime})/\kappa^{\prime})^{2}$
as $\kappa^{\prime}\to\kappa$. Thus we find
\begin{equation}
  |N_{\kappa\lambda}|=\sqrt{\frac{\kappa\sinh(\nu\pi)}{\nu}}.\label{eq:24}
\end{equation}
Second, the energy eigenvalues are determined by requiring that
$R_{\kappa\lambda}$ and $R_{\kappa^{\prime}\lambda}$ be orthogonal for
$\kappa\neq\kappa^{\prime}$; that is,
$\int_{0}^{\infty}\!dr\,\overline{R_{\kappa^{\prime}\lambda}(r)}R_{\kappa\lambda}(r)=0$
for $\kappa\neq\kappa^{\prime}$, which is attained if and only if
$\sin(\nu\log(\kappa/\kappa^{\prime}))=0$. Thus,
$\nu\log(\kappa/\kappa^{\prime})$ must be an integer multiple of
$\pi$:
\begin{equation}
  \nu\log\frac{\kappa}{\kappa^{\prime}}=-n\pi,\quad n\in\mathbb{Z},\label{eq:25}
\end{equation}
where the minus sign on the right hand side is just a convention. The
solution to this condition is given by
\begin{equation}
  \kappa_{n}=\kappa_{\ast}\exp\left(-\frac{n\pi}{\nu}\right),\label{eq:26}
\end{equation}
where $\kappa_{\ast}(>0)$ is an arbitrary reference scale with the
dimension of inverse length, which must be introduced on dimensional
grounds. Putting these together, we obtain an infinite number of
discrete negative energy eigenvalues:
\begin{equation}
  E_{n}=-\frac{\hbar^{2}\kappa_{\ast}^{2}}{2m}\exp\left(-\frac{2n\pi}{\nu}\right),\quad n\in\mathbb{Z}.\label{eq:27}
\end{equation}
These are the binding energies of the boundary-localized two-body
bound states. Note that the spatial extent of these bound states is
about $r\approx1/\kappa_{n}=\kappa_{\ast}^{-1}\e^{n\pi/\nu}$, which
follows from the asymptotic behavior
$R_{\kappa_{n}\lambda}(r)\to N_{\kappa_{n}\lambda}\e^{-\kappa_{n}r}$
as $r\to\infty$;\footnote{For more precise estimation, one should
  compute the expectation value of the distance
  $|x_{1}-x_{2}|=|r\sin\theta|$.} see the left panel of
Fig.~\ref{figure:4}. Note also that $E_{n}$ and
$R_{\kappa_{n}\lambda}(r)$ fulfill the relations
$E_{n-1}=E_{n}\e^{2\pi/\nu}$ and
$R_{\kappa_{n-1}\lambda}(r)=\e^{\pi/(2\nu)}R_{\kappa_{n}\lambda}(\e^{\pi/\nu}r)$,
which, through Eq.~\eqref{eq:10}, guarantees the scaling law
\eqref{eq:3} discussed in the introduction with the scaling factor
$\e^{t}=\e^{\pi/\nu}$.

\subsection{Two-body scattering off the boundary}
\label{section:3.3}
Let us finally consider the case $E>0$ in the channel
$\lambda=\lambda_{0}<0$. In this case, we are interested in the
solution to the radial equation \eqref{eq:11b} whose asymptotic
behavior as $r\to\infty$ is the linear combination of plane waves
$R_{k\lambda}(r)\to \e^{-ikr}+S_{\lambda}(k)\e^{ikr}$, where
$S_{\lambda}(k)$ is a linear combination coefficient and
$k=\sqrt{2mE/\hbar^{2}}>0$. The full solution is given by\footnote{For
  $E=\hbar^{2}k^{2}/(2m)$ and $\lambda=-\nu^{2}$, Eq.~\eqref{eq:11b}
  is equivalent to the Bessel differential equation
  $(r^{2}d^{2}/dr^{2}+rd/dr+k^{2}r^{2}+\nu^{2})(r^{-\frac{1}{2}}R)=0$. The
  two independent solutions to this equation are the Hankel functions
  of the first and second kind, $H^{(1)}_{i\nu}(\kappa r)$ and
  $H^{(2)}_{i\nu}(\kappa r)$, respectively. Note that the Hankel
  functions and the modified Bessel function of the second kind are
  related as
  $H^{(1)}_{i\nu}(x)=(2/(i\pi))\e^{-\frac{\nu\pi}{2}}K_{i\nu}(\e^{-\frac{i\pi}{2}}x)$
  and
  $H^{(2)}_{i\nu}(x)=(2/(i\pi))\e^{\frac{\nu\pi}{2}}K_{i\nu}(\e^{\frac{i\pi}{2}}x)$. For
  more details of the Hankel functions, see, e.g.,
  Ref.~\cite{Watson:1995}.}
\begin{align}
  R_{k\lambda}(r)=\sqrt{\frac{2kr}{\pi}}\left(\e^{\frac{i\pi}{4}}K_{i\nu}(\e^{\frac{i\pi}{2}}kr)+S_{\lambda}(k)\e^{-\frac{i\pi}{4}}K_{i\nu}(\e^{-\frac{i\pi}{2}}kr)\right).\label{eq:28}
\end{align}
Note that Eq.~\eqref{eq:28} together with Eq.~\eqref{eq:16} describes
the superposition of an incoming wave to $r=0$ and an outgoing wave
from $r=0$, both of which localize to $\theta=0$; that is, it
describes the two-body bound state scattered off the boundary, where
$S_{\lambda}(k)$ plays the role of the reflection amplitude (see the
right panel of Fig.~\ref{figure:4}). Note also that the scattering
solution \eqref{eq:28} is no longer localized to the boundary $r=0$.

It should be noted that at this stage $S_{\lambda}(k)$ is an arbitrary
constant. In order to determine $S_{\lambda}(k)$, we require that the
scattering solution \eqref{eq:28} be orthogonal to all the bound-state
solutions \eqref{eq:19}. (Note that the energy eigenfunctions should
be orthogonal if their eigenvalues are different.) In exactly the same
way as for Eq.~\eqref{eq:22}, one obtains the following relation:
\begin{align}
  (-\kappa_{n}^{2}-k^{2})\int_{0}^{\infty}\!\!\!dr\,\overline{R_{\kappa_{n}\lambda}(r)}R_{k\lambda}(r)
  &=-\lim_{r\to0}\left(\overline{R_{\kappa_{n}\lambda}(r)}R_{k\lambda}^{\prime}(r)-\overline{R_{\kappa_{n}\lambda}^{\prime}(r)}R_{k\lambda}(r)\right)\nonumber\\
  &=\frac{2\sqrt{\kappa_{n}k}}{\pi\sinh(\nu\pi)}\overline{N_{\kappa_{n}\lambda}}\biggl[\e^{\frac{i\pi}{4}}\sin\left(\nu\log\left(\frac{k}{\kappa_{n}}\right)+\frac{i\nu\pi}{2}\right)\nonumber\\
  &\quad
    +S_{\lambda}(k)\e^{-\frac{i\pi}{4}}\sin\left(\nu\log\left(\frac{k}{\kappa_{n}}\right)-\frac{i\nu\pi}{2}\right)\biggr].\label{eq:29}
\end{align}
Hence, in order to guarantee the orthogonality relation
$\int_{0}^{\infty}\!dr\,\overline{R_{\kappa_{n}\lambda}(r)}R_{k\lambda}(r)=0$
for any $k>0$ and $n\in\mathbb{Z}$, the coefficient $S_{\lambda}(k)$
must be of the following form:
\begin{equation}
  S_{\lambda}(k)=-\e^{\frac{i\pi}{2}}\frac{\sin\left(\nu\log\left(\frac{k}{\kappa_{\ast}}\right)+\frac{i\nu\pi}{2}\right)}{\sin\left(\nu\log\left(\frac{k}{\kappa_{\ast}}\right)-\frac{i\nu\pi}{2}\right)}.\label{eq:30}
\end{equation}
This is the reflection amplitude off the boundary for the two-body
bound state. This amplitude, which satisfies the unitarity condition
$\overline{S_{\lambda}(k)}S_{\lambda}(k)=1$, is a periodic function of
$\log k$ with the period $\pi/\nu$. This log-periodicity is a
manifestation of discrete scale invariance
$S_{\lambda}(\e^{n\pi/\nu}k)=S_{\lambda}(k)$ in the scattering
problem.\footnote{Once given the relation
  $S_{\lambda}(\e^{n\pi/\nu}k)=S_{\lambda}(k)$ for any
  $n\in\mathbb{Z}$, we can say that the reflection amplitude is of the
  form $S_{\lambda}(k)=f_{\lambda}(\log k)$, where $f_{\lambda}$ is a
  periodic function with the period $\pi/\nu$. In general, in
  scattering problems discrete scale invariance manifests itself in a
  periodic oscillation of the S-matrix as a function of $\log k$. For
  more details, see Ref.~\cite{Ohya:2021hju}.} We also note that
Eq.~\eqref{eq:30} has simple poles at
$k=i\kappa_{n}=\kappa_{\ast}\e^{i\pi/2-n\pi/\nu}$ in the complex
$k$-plane. In fact, it behaves as follows:
\begin{equation}
  S_{\lambda}(k)\to\frac{i|N_{\kappa_{n}\lambda}|^{2}}{k-i\kappa_{n}}+O(1)\quad\text{as $k\to i\kappa_{n}$}.\label{eq:31}
\end{equation}
These simple poles are the manifestation of the presence of infinitely
many bound states that satisfy the geometric scaling
$E_{n+1}/E_{n}=\kappa_{n+1}^{2}/\kappa_{n}^{2}=\e^{-2\pi/\nu}$.

We note in closing that the reflection amplitude \eqref{eq:30} can be
regarded as the scattering matrix (S-matrix) element
$S_{k\lambda,k^{\prime}\lambda^{\prime}}=(\Psi^{\text{out}}_{k\lambda},\Psi^{\text{in}}_{k^{\prime}\lambda^{\prime}})$
for $\lambda=\lambda^{\prime}=\lambda_{0}(<0)$, where
$\Psi^{\text{in}}_{k\lambda}=r^{-1/2}R_{k\lambda}\Theta_{\lambda}$ is
the in-state,
$\Psi^{\text{out}}_{k\lambda}=\overline{\Psi^{\text{in}}_{k\lambda}}$
is the out-state given by the complex conjugate (i.e., time reversal)
of the in-state, and $(\cdot,\cdot)$ is the inner product defined by
$(f,g)=\int_{0}^{\infty}\!\int_{0}^{\infty}\overline{f}g\,dx_{1}dx_{2}=\int_{0}^{\infty}\!\int_{-\pi/4}^{\pi/4}\overline{f}g\,rdrd\theta$. In
fact, it follows from the orthonormality
$\int_{-\pi/4}^{\pi/4}\Theta_{\lambda}\Theta_{\lambda^{\prime}}d\theta=\delta_{\lambda\lambda^{\prime}}$,\footnote{Without
  any loss of generality, the angular wavefunction $\Theta_{\lambda}$
  can be chosen to be real for any
  $\lambda\in\{\lambda_{0},\lambda_{1},\cdots\}$.} the identity
$(k^{2}-k^{\prime2})R_{k\lambda}R_{k^{\prime}\lambda}=(d/dr)(R_{k\lambda}R_{k^{\prime}\lambda}^{\prime}-R_{k\lambda}^{\prime}R_{k^{\prime}\lambda})$,
and the asymptotic behavior
$R_{k\lambda}\to \e^{-ikr}+S_{\lambda}(k)\e^{ikr}$ that this S-matrix
element takes the form
$S_{k\lambda,k^{\prime}\lambda^{\prime}}=2\pi\delta(k-k^{\prime})\delta_{\lambda\lambda^{\prime}}S_{\lambda}(k)$. One
nice thing in this formulation is that it is obvious that there is no
scattering between different channels $\lambda$ and
$\lambda^{\prime}$.

\pagebreak[4]
\section{Conclusion}
\label{section:4}
In this paper, we have introduced a toy scale-invariant model of two
identical bosons on the half-line $\mathbb{R}_{+}$, where
interparticle interaction is described by the pairwise
$\delta$-function potential with the particular position-dependent
coupling strength given by Eq.~\eqref{eq:2}. We have seen that, if the
two-body interaction is sufficiently attractive, continuous scale
invariance is broken down to discrete scale invariance. In the
bound-state problem where the bosons are bound together and localized
to the boundary, this discrete scale invariance manifests itself in
the onset of the geometric sequence of binding energies. In the
scattering problem where the two-body bound state is scattered by the
boundary, on the other hand, this discrete scale invariance manifests
itself in the log-periodic behavior of the reflection
amplitude. Hence, by breaking translation invariance of this
one-dimensional problem, we can construct a two-body model that
exhibits the Efimov effect. In contrast to the ordinary Efimov effect
in three-body problems in three dimensions, our model can be solved
exactly by just using undergraduate-level calculus.

Finally, it should be mentioned the stability issue of the model and
its cure. As is evident from Eq.~\eqref{eq:27}, there is no lower
bound in the energy spectrum $\{E_{n}\}$ for $g_{0}<g_{\ast}$. This
absence of ground state is inevitable if the system is invariant under
the full discrete scale invariance that forms the group
$\mathbb{Z}$. (As discussed in the introduction, the full discrete
scale invariance leads to the geometric sequence
$\{E_{0},E_{0}\e^{\pm2t_{\ast}},E_{0}\e^{\pm4t_{\ast}},\cdots\}$,
which cannot be bounded from below if $E_{0}<0$.) In order to make the
spectrum lower-bounded, we therefore have to break this invariance
under $\mathbb{Z}$. The easiest way to do this is to replace the
short-distance singularity of the inverse-square potential by, e.g., a
square-well potential. Such regularization procedures have been widely
studied over the years in the context of renormalization of the
inverse-square potential. For more details, we refer to
Refs.~\cite{Camblong:2000ec,Camblong:2000ax,Coon:2002sua,Bawin:2003dm,Braaten:2004pg,Hammer:2005sa,Essin:2006,Kaplan:2009kr}.

\begin{appendices}
  \setcounter{equation}{0}
  \renewcommand{\theequation}{\thesection.\arabic{equation}}
  \section{Modified Bessel functions of imaginary order}
  \label{appendix:A}
  In this section, we summarize the short- and long-distance behaviors
  of the modified Bessel functions. For details, we refer to
  Ref.~\cite{Watson:1995}.

  First of all, the modified Bessel function of the second kind with
  imaginary order is defined as follows:
  \begin{equation}
    K_{i\nu}(z)=\frac{i\pi}{2}\frac{I_{i\nu}(z)-I_{-i\nu}(z)}{\sinh(\nu\pi)},\quad\nu\in\mathbb{R}\setminus\{0\},\label{eq:A1}
  \end{equation}
  where $I_{i\nu}$ is the modified Bessel function of the first kind
  given by the following series:
  \begin{equation}
    I_{i\nu}(z)=\e^{i\nu\log\frac{z}{2}}\sum_{n=0}^{\infty}\frac{1}{n!\Gamma(1+n+i\nu)}\left(\frac{z}{2}\right)^{2n}.\label{eq:A2}
  \end{equation}
  Here $\Gamma$ is the gamma function. It follows immediately from the
  definition \eqref{eq:A1} that $K_{-i\nu}(z)=K_{i\nu}(z)$. It also
  follows from Eqs.~\eqref{eq:A1} and \eqref{eq:A2} that
  $\overline{K_{i\nu}(z)}=K_{i\nu}(z)$ for $z>0$.

  The short-distance behavior of $K_{i\nu}$ is governed by the $n=0$
  term in Eq.~\eqref{eq:A2}. By using the polar form of the gamma
  function,
  \begin{align}
    \Gamma(1+i\nu)
    &=|\Gamma(1+i\nu)|\e^{i\arg\Gamma(1+i\nu)}\nonumber\\
    &=\sqrt{\frac{\nu\pi}{\sinh(\nu\pi)}}\e^{i\arg\Gamma(1+i\nu)},\label{eq:A3}
  \end{align}
  where $\arg\Gamma(1+i\nu)$ stands for the argument of
  $\Gamma(1+i\nu)$, we see that $K_{i\nu}(z)$ behaves as follows:
  \begin{align}
    K_{i\nu}(z)\to-\sqrt{\frac{\pi}{\nu\sinh(\nu\pi)}}\sin\left(\nu\log\frac{z}{2}-\arg\Gamma(1+i\nu)\right)+O(z^{2})\quad\text{as $|z|\to0$}.\label{eq:A4}
  \end{align}

  The long-distance behavior, on the other hand, is known to be of the
  following form:
  \begin{equation}
    K_{i\nu}(z)\to\sqrt{\frac{\pi}{2z}}\e^{-z}\left[1+O(\tfrac{1}{z})\right]\quad\text{as $|z|\to\infty$}.\label{eq:A5}
  \end{equation}

  \setcounter{equation}{0}
  \section{Boundary as the infinitely heavy third particle}
  \label{appendix:B}
  In this section we show that the two-body problem on the half-line
  $\mathbb{R}_{+}$ discussed in the main text is equivalent to a
  three-body problem on the whole line $\mathbb{R}$ with an infinitely
  heavy third particle. We note that this section is not necessary for
  understanding the main text.

  To begin with, let us first define some notation. Let
  $z_{1},z_{2}\in\mathbb{R}$ be the coordinates of two identical
  bosons of mass $m$ and $z_{3}\in\mathbb{R}$ be the coordinate of a
  third particle of mass $M$. The Hamiltonian of these particles is
  assumed to be of the following form:
  \begin{align}
    H_{\text{3-body}}=-\frac{\hbar^{2}}{2m}\left(\frac{\partial^{2}}{\partial z_{1}^{2}}+\frac{\partial^{2}}{\partial z_{2}^{2}}\right)-\frac{\hbar^{2}}{2M}\frac{\partial^{2}}{\partial z_{3}^{2}}+g(z_{1}-z_{3})\delta(z_{1}-z_{2})+U(z_{1}-z_{3})+U(z_{2}-z_{3}),\label{eq:B1}
  \end{align}
  where $g(x)=\hbar^{2}g_{0}/(m|x|)$ is the position-dependent
  coupling strength that makes the two-body interaction between the
  identical bosons scale invariant. Note that the two bosons interact
  only when they are in the same position, and that the interaction
  strength depends on their position relative to the massive third
  particle. $U$ is a pairwise interaction between one of the identical
  bosons and the third particle whose explicit form is irrelevant for
  the moment. Note that Eq.~\eqref{eq:B1} is invariant under the
  translation $(z_{1},z_{2},z_{3})\mapsto(z_{1}+a,z_{2}+a,z_{3}+a)$
  for any $a\in\mathbb{R}$. Hence the total momentum is conserved and
  the center-of-mass motion is that of a free particle. Below, we will
  show that, by taking the limit $M\to\infty$, the relative
  Hamiltonian that describes the relative motion of this three-body
  system can be reduced to the two-body Hamiltonian on the half-line
  discussed in the main text.

  Let us first separate the center-of-mass Hamiltonian from
  Eq.~\eqref{eq:B1} and identify the relative Hamiltonian. To this
  end, it is convenient to introduce the Jacobi coordinates
  $(y_{1},y_{2},y_{3})$ as follows:
  \begin{subequations}
    \begin{align}
      y_{1}&=z_{1}-z_{2},\label{eq:B2a}\\
      y_{2}&=\frac{z_{1}+z_{2}}{2}-z_{3},\label{eq:B2b}\\
      y_{3}&=\frac{mz_{1}+mz_{2}+Mz_{3}}{2m+M}.\label{eq:B2c}
    \end{align}
  \end{subequations}
  Physically, $y_{1}$ is the relative coordinate for the identical
  bosons; $y_{2}$ is the relative coordinate for the center-of-mass of
  the identical bosons and the heavy particle; and $y_{3}$ is the
  center-of-mass coordinate for the three particles. A main advantage
  in this coordinate system is the following equality:
  \begin{equation}
    \frac{1}{m}\frac{\partial^{2}}{\partial z_{1}^{2}}+\frac{1}{m}\frac{\partial^{2}}{\partial z_{2}^{2}}+\frac{1}{M}\frac{\partial^{2}}{\partial z_{3}^{2}}
    =\frac{1}{\mu_{1}}\frac{\partial^{2}}{\partial y_{1}^{2}}+\frac{1}{\mu_{2}}\frac{\partial^{2}}{\partial y_{2}^{2}}+\frac{1}{\mu_{3}}\frac{\partial^{2}}{\partial y_{3}^{2}},\label{eq:B3}
  \end{equation}
  where
  \begin{subequations}
    \begin{align}
      \mu_{1}&=\left(\frac{1}{m}+\frac{1}{m}\right)^{-1}=\frac{m}{2},\label{eq:B4a}\\
      \mu_{2}&=\left(\frac{1}{2m}+\frac{1}{M}\right)^{-1}=\frac{2mM}{2m+M},\label{eq:B4b}\\
      \mu_{3}&=2m+M.\label{eq:B4c}
    \end{align}
  \end{subequations}
  Physically, $\mu_{1}$ is the reduced mass for the identical bosons;
  $\mu_{2}$ is the reduced mass for the center-of-mass of the
  identical bosons and the heavy particle; and $\mu_{3}$ is the total
  mass of the three particles. It is now straightforward to show that
  the three-body Hamiltonian \eqref{eq:B1} can be written as
  $H_{\text{3-body}}=H_{\text{cm}}+H_{\text{rel}}$, where
  \begin{subequations}
    \begin{align}
      H_{\text{cm}}&=-\frac{\hbar^{2}}{2\mu_{3}}\frac{\partial^{2}}{\partial y_{3}^{2}},\label{eq:B5a}\\
      H_{\text{rel}}&=-\frac{\hbar^{2}}{2\mu_{1}}\frac{\partial^{2}}{\partial y_{1}^{2}}-\frac{\hbar^{2}}{2\mu_{2}}\frac{\partial^{2}}{\partial y_{2}^{2}}+g(y_{2}+\tfrac{1}{2}y_{1})\delta(y_{1})+U(y_{2}+\tfrac{1}{2}y_{1})+U(y_{2}-\tfrac{1}{2}y_{1}).\label{eq:B5b}
    \end{align}
  \end{subequations}
  $H_{\text{cm}}$ is the center-of-mass Hamiltonian so that
  $H_{\text{rel}}=H_{\text{3-body}}-H_{\text{cm}}$ describes the
  relative motion and internal energy of the three-body system. In the
  following, we focus on $H_{\text{rel}}$.

  Now let us consider the situation where the third particle is much
  heavier than the identical bosons, $m/M\ll1$. In the extreme case
  $M\to\infty$, where $\mu_{2}\to2m$, Eq.~\eqref{eq:B5b} reduces to
  \begin{align}
    H_{\text{eff}}
    &=\lim_{M\to\infty}H_{\text{rel}}\nonumber\\
    &=-\frac{\hbar^{2}}{m}\frac{\partial^{2}}{\partial y_{1}^{2}}-\frac{\hbar^{2}}{4m}\frac{\partial^{2}}{\partial y_{2}^{2}}+g(y_{2}+\tfrac{1}{2}y_{1})\delta(y_{1})+U(y_{2}+\tfrac{1}{2}y_{1})+U(y_{2}-\tfrac{1}{2}y_{1}).\label{eq:B6}
  \end{align}
  To standardize the expression, let us introduce a new coordinate
  system $(x_{1},x_{2})$ defined by
  \begin{subequations}
    \begin{align}
      x_{1}&=y_{2}+\frac{1}{2}y_{1}=z_{1}-z_{3},\label{eq:B7a}\\
      x_{2}&=y_{2}-\frac{1}{2}y_{1}=z_{2}-z_{3},\label{eq:B7b}
    \end{align}
  \end{subequations}
  in which Eq.~\eqref{eq:B6} takes the following form:
  \begin{align}
    H_{\text{eff}}=-\frac{\hbar^{2}}{2m}\left(\frac{\partial^{2}}{\partial x_{1}^{2}}+\frac{\partial^{2}}{\partial x_{2}^{2}}\right)+g(x_{1})\delta(x_{1}-x_{2})+U(x_{1})+U(x_{2}).\label{eq:B8}
  \end{align}
  This Hamiltonian describes the relative motion of the three-body
  system in the center-of-mass frame and in the $M\to\infty$
  limit. Notice that $y_{3}\to z_{3}$ as $M\to\infty$ so that the
  infinitely heavy third particle remains at the origin in the
  center-of-mass frame. In this limit, the three-body system is
  described by the two-body Hamiltonian \eqref{eq:B8}.

  Finally, let us consider the confinement of the identical bosons on
  the half-line $\mathbb{R}_{+}$ by specifying $U$. One way to achieve
  this is to choose $U(x)=0$ for $x>0$ and $U(x)=\infty$ for $x<0$. An
  alternative way is to use the strong-coupling limit of the following
  $\delta$-function potential:
  \begin{align}
    U(x)=\frac{\hbar^{2}}{m}\gamma\delta(x),\label{eq:B9}
  \end{align}
  where $\gamma$ is a coupling constant. As discussed in
  Sec.~\ref{section:3.1}, $U(x_{1})$ is equivalent to the following
  connection conditions for the two-body wavefunction
  $\psi(x_{1},x_{2})$:
  \begin{subequations}
    \begin{align}
      -\frac{\partial\psi}{\partial x_{1}}(0_{+},x_{2})+\frac{\partial\psi}{\partial x_{1}}(0_{-},x_{2})+\gamma\left(\psi(0_{+},x_{2})+\psi(0_{-},x_{2})\right)=0,\label{eq:B10a}\\
      \psi(0_{+},x_{2})=\psi(0_{-},x_{2}),\label{eq:B10b}
    \end{align}
  \end{subequations}
  where $x_{2}\neq0$. In the strong-coupling limit $\gamma\to\infty$,
  Eq.~\eqref{eq:B10a} reduces to
  $\psi(0_{+},x_{2})+\psi(0_{-},x_{2})=0$, which, together with
  Eq.~\eqref{eq:B10b}, leads to the Dirichlet boundary condition
  $\psi(0_{+},x_{2})=\psi(0_{-},x_{2})=0$ that corresponds to
  $\Theta(\pi/4)=0$ in Eq.~\eqref{eq:15}. Similarly, for $U(x_{2})$,
  one can obtain $\psi(x_{1},0_{+})=\psi(x_{1},0_{-})=0$
  ($x_{1}\neq0$) that corresponds to $\Theta(-\pi/4)=0$. Notice that,
  under the Dirichlet boundary conditions, the two regions $x>0$ and
  $x<0$ are physically disconnected because the probability current
  density vanishes at $x=0$ and hence there is no probability current
  flow across the origin. Alternatively, one can say that particles
  cannot penetrate through the origin because the transmission
  amplitude for the $\delta$-function potential vanishes in the
  strong-coupling limit. Hence, if the identical bosons are initially
  on the positive half-line $\mathbb{R}_{+}$, they remain on this
  region forever. This is the two-body problem on the half-line
  discussed in the main text. Note that the effective two-body
  Hamiltonian \eqref{eq:B8} is no longer invariant under the
  translation $(x_{1},x_{2})\mapsto(x_{1}+a,x_{2}+a)$.
\end{appendices}

\printbibliography
\end{document}

%% file: figure1.eepic
%% Generated from figure1.xp on Fri Sep 16 13:10:44 JST 2022 by
%% ePiX-1.2.18
%% 
%%   Cartesian bounding box: [-1.3,6.3] x [-2.5,2.5]
%%   Actual size: 6.08 x 4cm
%%   Figure offset: left by 0cm, down by 0cm
%% 
%% usepackages epic,eepic,xcolor
%% 
\xdefinecolor{rgb_000000}{rgb}{0,0,0}%
\xdefinecolor{rgb_0000ff}{rgb}{0,0,1}%
\setlength{\unitlength}{1cm}%
\begin{picture}(6.08,4)(0,0)%
\path(1.04,2)(5.04,2)
\path(4.88184,1.96046)(4.92138,1.96046)
\path(4.88184,1.97364)(4.96092,1.97364)
\path(4.88184,1.98682)(5.00046,1.98682)
\path(4.88184,2)(5.04,2)
\path(4.88184,2.01318)(5.00046,2.01318)
\path(4.88184,2.02636)(4.96092,2.02636)
\path(4.88184,2.03954)(4.92138,2.03954)
\path(5.02682,1.99561)(5.02682,2.00439)
\path(5.01364,1.99121)(5.01364,2.00879)
\path(5.00046,1.98682)(5.00046,2.01318)
\path(4.98728,1.98243)(4.98728,2.01757)
\path(4.9741,1.97803)(4.9741,2.02197)
\path(4.96092,1.97364)(4.96092,2.02636)
\path(4.94774,1.96925)(4.94774,2.03075)
\path(4.93456,1.96485)(4.93456,2.03515)
\path(4.92138,1.96046)(4.92138,2.03954)
\path(4.9082,1.95607)(4.9082,2.04393)
\path(4.89502,1.95167)(4.89502,2.04833)
\path(4.88184,1.94728)(4.88184,2.05272)
\path(4.88184,2)(4.88184,1.94728)(5.04,2)(4.88184,2.05272)(4.88184,2)
\path(1.04,0)(1.04,4)
\path(1.08722,3.84184)(1.03341,3.98023)
\path(1.07294,3.84184)(1.02682,3.96046)
\path(1.05866,3.84184)(1.02023,3.94069)
\path(1.04439,3.84184)(1.01364,3.92092)
\path(1.03011,3.84184)(1.00705,3.90115)
\path(1.01583,3.84184)(1.00046,3.88138)
\path(1.00156,3.84184)(0.993871,3.86161)
\path(1.04439,3.98682)(1.0343,3.9829)
\path(1.04879,3.97364)(1.0286,3.96579)
\path(1.05318,3.96046)(1.0229,3.94869)
\path(1.05757,3.94728)(1.01719,3.93158)
\path(1.06197,3.9341)(1.01149,3.91448)
\path(1.06636,3.92092)(1.00579,3.89737)
\path(1.07075,3.90774)(1.00009,3.88027)
\path(1.07515,3.89456)(0.994388,3.86316)
\path(1.07954,3.88138)(0.988686,3.84606)
\path(1.08393,3.8682)(1.01614,3.84184)
\path(1.08833,3.85502)(1.05443,3.84184)
\path(1.04,3.84184)(1.09272,3.84184)(1.04,4)(0.987281,3.84184)(1.04,3.84184)
\color{rgb_0000ff}%
\path(1.36,4)(1.38,3.88235)(1.4,3.77778)(1.42,3.68421)(1.44,3.6)
  (1.46,3.52381)(1.48,3.45455)(1.5,3.3913)(1.52,3.33333)(1.54,3.28)
  (1.56,3.23077)(1.58,3.18519)(1.6,3.14286)(1.62,3.10345)
  (1.64,3.06667)(1.66,3.03226)(1.68,3)(1.7,2.9697)(1.72,2.94118)
  (1.74,2.91429)(1.76,2.88889)(1.78,2.86486)(1.8,2.84211)
  (1.82,2.82051)(1.84,2.8)(1.86,2.78049)(1.88,2.7619)(1.9,2.74419)
  (1.92,2.72727)(1.94,2.71111)(1.96,2.69565)(1.98,2.68085)(2,2.66667)
  (2.02,2.65306)(2.04,2.64)(2.06,2.62745)(2.08,2.61538)(2.1,2.60377)
  (2.12,2.59259)(2.14,2.58182)(2.16,2.57143)(2.18,2.5614)
  (2.2,2.55172)(2.22,2.54237)(2.24,2.53333)(2.26,2.52459)
  (2.28,2.51613)(2.3,2.50794)(2.32,2.5)(2.34,2.49231)(2.36,2.48485)
  (2.38,2.47761)(2.4,2.47059)(2.42,2.46377)(2.44,2.45714)
  (2.46,2.4507)(2.48,2.44444)(2.5,2.43836)(2.52,2.43243)
  (2.54,2.42667)(2.56,2.42105)(2.58,2.41558)(2.6,2.41026)
  (2.62,2.40506)(2.64,2.4)(2.66,2.39506)(2.68,2.39024)(2.7,2.38554)
  (2.72,2.38095)(2.74,2.37647)(2.76,2.37209)(2.78,2.36782)
  (2.8,2.36364)(2.82,2.35955)(2.84,2.35556)(2.86,2.35165)
  (2.88,2.34783)(2.9,2.34409)(2.92,2.34043)(2.94,2.33684)
  (2.96,2.33333)(2.98,2.3299)(3,2.32653)(3.02,2.32323)(3.04,2.32)
  (3.06,2.31683)(3.08,2.31373)(3.1,2.31068)(3.12,2.30769)
  (3.14,2.30476)(3.16,2.30189)(3.18,2.29907)(3.2,2.2963)
  (3.22,2.29358)(3.24,2.29091)(3.26,2.28829)(3.28,2.28571)
  (3.3,2.28319)(3.32,2.2807)(3.34,2.27826)(3.36,2.27586)(3.38,2.2735)
  (3.4,2.27119)(3.42,2.26891)(3.44,2.26667)(3.46,2.26446)
  (3.48,2.2623)(3.5,2.26016)(3.52,2.25806)(3.54,2.256)(3.56,2.25397)
  (3.58,2.25197)(3.6,2.25)(3.62,2.24806)(3.64,2.24615)(3.66,2.24427)
  (3.68,2.24242)(3.7,2.2406)(3.72,2.23881)(3.74,2.23704)
  (3.76,2.23529)(3.78,2.23358)(3.8,2.23188)(3.82,2.23022)
  (3.84,2.22857)(3.86,2.22695)(3.88,2.22535)(3.9,2.22378)
  (3.92,2.22222)(3.94,2.22069)(3.96,2.21918)(3.98,2.21769)(4,2.21622)
  (4.02,2.21477)(4.04,2.21333)(4.06,2.21192)(4.08,2.21053)
  (4.1,2.20915)(4.12,2.20779)(4.14,2.20645)(4.16,2.20513)
  (4.18,2.20382)(4.2,2.20253)(4.22,2.20126)(4.24,2.2)(4.26,2.19876)
  (4.28,2.19753)(4.3,2.19632)(4.32,2.19512)(4.34,2.19394)
  (4.36,2.19277)(4.38,2.19162)(4.4,2.19048)(4.42,2.18935)
  (4.44,2.18824)(4.46,2.18713)(4.48,2.18605)(4.5,2.18497)
  (4.52,2.18391)(4.54,2.18286)(4.56,2.18182)(4.58,2.18079)
  (4.6,2.17978)(4.62,2.17877)(4.64,2.17778)(4.66,2.1768)
  (4.68,2.17582)(4.7,2.17486)(4.72,2.17391)(4.74,2.17297)
  (4.76,2.17204)(4.78,2.17112)(4.8,2.17021)(4.82,2.16931)
  (4.84,2.16842)(4.86,2.16754)(4.88,2.16667)(4.9,2.1658)
  (4.92,2.16495)(4.94,2.1641)(4.96,2.16327)(4.98,2.16244)(5,2.16162)
  (5.02,2.1608)(5.04,2.16)
\path(1.36,0)(1.38,0.117647)(1.4,0.222222)(1.42,0.315789)(1.44,0.4)
  (1.46,0.47619)(1.48,0.545455)(1.5,0.608696)(1.52,0.666667)
  (1.54,0.72)(1.56,0.769231)(1.58,0.814815)(1.6,0.857143)
  (1.62,0.896552)(1.64,0.933333)(1.66,0.967742)(1.68,1)(1.7,1.0303)
  (1.72,1.05882)(1.74,1.08571)(1.76,1.11111)(1.78,1.13514)
  (1.8,1.15789)(1.82,1.17949)(1.84,1.2)(1.86,1.21951)(1.88,1.2381)
  (1.9,1.25581)(1.92,1.27273)(1.94,1.28889)(1.96,1.30435)
  (1.98,1.31915)(2,1.33333)(2.02,1.34694)(2.04,1.36)(2.06,1.37255)
  (2.08,1.38462)(2.1,1.39623)(2.12,1.40741)(2.14,1.41818)
  (2.16,1.42857)(2.18,1.4386)(2.2,1.44828)(2.22,1.45763)
  (2.24,1.46667)(2.26,1.47541)(2.28,1.48387)(2.3,1.49206)(2.32,1.5)
  (2.34,1.50769)(2.36,1.51515)(2.38,1.52239)(2.4,1.52941)
  (2.42,1.53623)(2.44,1.54286)(2.46,1.5493)(2.48,1.55556)
  (2.5,1.56164)(2.52,1.56757)(2.54,1.57333)(2.56,1.57895)
  (2.58,1.58442)(2.6,1.58974)(2.62,1.59494)(2.64,1.6)(2.66,1.60494)
  (2.68,1.60976)(2.7,1.61446)(2.72,1.61905)(2.74,1.62353)
  (2.76,1.62791)(2.78,1.63218)(2.8,1.63636)(2.82,1.64045)
  (2.84,1.64444)(2.86,1.64835)(2.88,1.65217)(2.9,1.65591)
  (2.92,1.65957)(2.94,1.66316)(2.96,1.66667)(2.98,1.6701)(3,1.67347)
  (3.02,1.67677)(3.04,1.68)(3.06,1.68317)(3.08,1.68627)(3.1,1.68932)
  (3.12,1.69231)(3.14,1.69524)(3.16,1.69811)(3.18,1.70093)
  (3.2,1.7037)(3.22,1.70642)(3.24,1.70909)(3.26,1.71171)
  (3.28,1.71429)(3.3,1.71681)(3.32,1.7193)(3.34,1.72174)
  (3.36,1.72414)(3.38,1.7265)(3.4,1.72881)(3.42,1.73109)
  (3.44,1.73333)(3.46,1.73554)(3.48,1.7377)(3.5,1.73984)
  (3.52,1.74194)(3.54,1.744)(3.56,1.74603)(3.58,1.74803)(3.6,1.75)
  (3.62,1.75194)(3.64,1.75385)(3.66,1.75573)(3.68,1.75758)
  (3.7,1.7594)(3.72,1.76119)(3.74,1.76296)(3.76,1.76471)
  (3.78,1.76642)(3.8,1.76812)(3.82,1.76978)(3.84,1.77143)
  (3.86,1.77305)(3.88,1.77465)(3.9,1.77622)(3.92,1.77778)
  (3.94,1.77931)(3.96,1.78082)(3.98,1.78231)(4,1.78378)(4.02,1.78523)
  (4.04,1.78667)(4.06,1.78808)(4.08,1.78947)(4.1,1.79085)
  (4.12,1.79221)(4.14,1.79355)(4.16,1.79487)(4.18,1.79618)
  (4.2,1.79747)(4.22,1.79874)(4.24,1.8)(4.26,1.80124)(4.28,1.80247)
  (4.3,1.80368)(4.32,1.80488)(4.34,1.80606)(4.36,1.80723)
  (4.38,1.80838)(4.4,1.80952)(4.42,1.81065)(4.44,1.81176)
  (4.46,1.81287)(4.48,1.81395)(4.5,1.81503)(4.52,1.81609)
  (4.54,1.81714)(4.56,1.81818)(4.58,1.81921)(4.6,1.82022)
  (4.62,1.82123)(4.64,1.82222)(4.66,1.8232)(4.68,1.82418)
  (4.7,1.82514)(4.72,1.82609)(4.74,1.82703)(4.76,1.82796)
  (4.78,1.82888)(4.8,1.82979)(4.82,1.83069)(4.84,1.83158)
  (4.86,1.83246)(4.88,1.83333)(4.9,1.8342)(4.92,1.83505)(4.94,1.8359)
  (4.96,1.83673)(4.98,1.83756)(5,1.83838)(5.02,1.8392)(5.04,1.84)
\put(0.969708,2){\makebox(0,0)[r]{\hbox{\color{rgb_000000}\scriptsize $0$}}}
\put(5.11029,2){\makebox(0,0)[l]{\hbox{\color{rgb_000000}\scriptsize $x$}}}
\put(0.934562,4){\makebox(0,0)[tr]{\hbox{\color{rgb_000000}\scriptsize $g$}}}
\put(1.59029,3.40363){\makebox(0,0)[bl]{\hbox{\color{rgb_000000}\scriptsize \textcolor{blue}{Case $g_{0}>0$}}}}
\put(1.59029,0.596375){\makebox(0,0)[tl]{\hbox{\color{rgb_000000}\scriptsize \textcolor{blue}{Case $g_{0}<0$}}}}
\end{picture}%

%% file: figure2.eepic
%% Generated from figure2.xp on Fri Sep 16 13:10:45 JST 2022 by
%% ePiX-1.2.18
%% 
%%   Cartesian bounding box: [-0.04,1.08] x [-0.05,1.07]
%%   Actual size: 4 x 4cm
%%   Figure offset: left by 0cm, down by 0cm
%% 
%% usepackages epic,eepic,xcolor
%% 
\xdefinecolor{rgb_000000}{rgb}{0,0,0}%
\xdefinecolor{rgb_0000ff}{rgb}{0,0,1}%
\xdefinecolor{rgb_b2b2b2}{rgb}{0.698039,0.698039,0.698039}%
\xdefinecolor{rgb_b2b2ff}{rgb}{0.698039,0.698039,1}%
\xdefinecolor{rgb_cc0000}{rgb}{0.8,0,0}%
\setlength{\unitlength}{1cm}%
\begin{picture}(4,4)(0,0)%
\path(1.40555,1.44126)(1.40058,1.44621)(1.39559,1.45114)
  (1.39058,1.45605)(1.38556,1.46094)(1.38051,1.46581)
  (1.37545,1.47066)(1.37037,1.47549)(1.36526,1.4803)(1.36014,1.48509)
  (1.355,1.48986)(1.34984,1.49461)(1.34467,1.49934)(1.33947,1.50405)
  (1.33425,1.50874)(1.32902,1.51341)(1.32377,1.51806)(1.3185,1.52268)
  (1.31321,1.52729)(1.30791,1.53187)(1.30259,1.53644)
  (1.29724,1.54098)(1.29189,1.54551)(1.28651,1.55001)
  (1.28111,1.55449)(1.2757,1.55895)(1.27027,1.56339)(1.26483,1.5678)
  (1.25936,1.5722)(1.25388,1.57657)(1.24838,1.58092)(1.24287,1.58525)
  (1.23733,1.58956)(1.23178,1.59385)(1.22622,1.59811)
  (1.22064,1.60236)(1.21504,1.60658)(1.20942,1.61078)
  (1.20379,1.61496)(1.19814,1.61911)(1.19247,1.62324)
  (1.18679,1.62736)(1.18109,1.63144)(1.17538,1.63551)
  (1.16965,1.63955)(1.16391,1.64357)(1.15815,1.64757)
  (1.15237,1.65155)(1.14658,1.6555)(1.14077,1.65943)(1.13495,1.66334)
  (1.12911,1.66722)(1.12326,1.67108)(1.11739,1.67492)(1.1115,1.67874)
  (1.1056,1.68253)(1.09969,1.6863)(1.09376,1.69005)(1.08782,1.69377)
  (1.08186,1.69747)(1.07589,1.70114)(1.0699,1.7048)(1.0639,1.70842)
  (1.05789,1.71203)(1.05186,1.71561)(1.04582,1.71917)(1.03976,1.7227)
  (1.03369,1.72621)(1.0276,1.7297)(1.02151,1.73316)(1.0154,1.7366)
  (1.00927,1.74001)(1.00313,1.7434)(0.99698,1.74677)
  (0.990815,1.75011)(0.984637,1.75343)(0.978446,1.75672)
  (0.972242,1.75999)(0.966026,1.76324)(0.959796,1.76646)
  (0.953554,1.76965)(0.9473,1.77283)(0.941033,1.77597)
  (0.934754,1.77909)(0.928463,1.78219)(0.922159,1.78526)
  (0.915844,1.78831)(0.909517,1.79134)(0.903177,1.79433)
  (0.896826,1.79731)(0.890464,1.80026)(0.88409,1.80318)
  (0.877704,1.80608)(0.871307,1.80895)(0.864899,1.8118)
  (0.85848,1.81462)(0.85205,1.81742)(0.845609,1.82019)
  (0.839157,1.82294)(0.832694,1.82566)(0.82622,1.82836)
  (0.819736,1.83103)(0.813242,1.83367)(0.806737,1.83629)
  (0.800222,1.83889)(0.793697,1.84146)(0.787162,1.844)
  (0.780617,1.84652)(0.774062,1.84901)(0.767498,1.85147)
  (0.760923,1.85391)(0.75434,1.85633)(0.747746,1.85872)
  (0.741144,1.86108)(0.734532,1.86341)(0.727911,1.86573)
  (0.721281,1.86801)(0.714642,1.87027)(0.707995,1.8725)
  (0.701338,1.87471)(0.694673,1.87689)
\path(0.747746,1.85872)(0.721281,1.86801)
\path(0.836567,1.77863)(0.85752,1.77863)
\path(0.820097,1.7914)(0.862004,1.7914)
\path(0.803628,1.80417)(0.866488,1.80417)
\path(0.787159,1.81694)(0.870972,1.81694)
\path(0.770689,1.82971)(0.875455,1.82971)
\path(0.75422,1.84247)(0.879939,1.84247)
\path(0.737751,1.85524)(0.884423,1.85524)
\path(0.874081,1.82579)(0.874081,1.86557)
\path(0.86019,1.78624)(0.86019,1.86579)
\path(0.846299,1.77109)(0.846299,1.86601)
\path(0.832408,1.78186)(0.832408,1.86623)
\path(0.818517,1.79263)(0.818517,1.86646)
\path(0.804626,1.8034)(0.804626,1.86668)
\path(0.790736,1.81416)(0.790736,1.8669)
\path(0.776845,1.82493)(0.776845,1.86712)
\path(0.762954,1.8357)(0.762954,1.86734)
\path(0.749063,1.84647)(0.749063,1.86757)
\path(0.735172,1.85724)(0.735172,1.86779)
\path(0.870504,1.81561)(0.887972,1.86535)(0.721281,1.86801)
  (0.853036,1.76587)(0.870504,1.81561)
\color{rgb_0000ff}%
\path(0.142857,0.178571)(1.02576,2.89588)
\path(1.02604,2.85983)(1.01676,2.88368)
\path(1.02631,2.82378)(1.00776,2.87149)
\path(1.02658,2.78773)(0.99876,2.8593)
\path(1.02686,2.75168)(0.989759,2.8471)
\path(1.02112,2.73109)(0.980757,2.83491)
\path(1.00539,2.7362)(0.971756,2.82272)
\path(0.989666,2.74131)(0.962755,2.81052)
\path(0.973937,2.74642)(0.953754,2.79833)
\path(0.958209,2.75153)(0.944753,2.78614)
\path(0.94248,2.75664)(0.935752,2.77394)
\path(1.02587,2.88198)(1.01134,2.87633)
\path(1.02597,2.86809)(0.996911,2.85679)
\path(1.02608,2.8542)(0.982484,2.83725)
\path(1.02618,2.84031)(0.968058,2.81771)
\path(1.02629,2.82641)(0.953632,2.79816)
\path(1.0264,2.81252)(0.939206,2.77862)
\path(1.0265,2.79863)(0.929418,2.76088)
\path(1.02661,2.78474)(0.94894,2.75454)
\path(1.02671,2.77085)(0.968462,2.7482)
\path(1.02682,2.75695)(0.987984,2.74185)
\path(1.02692,2.74306)(1.00751,2.73551)
\path(0.97689,2.74546)(1.02703,2.72917)(1.02576,2.89588)
  (0.926751,2.76175)(0.97689,2.74546)
\color{rgb_b2b2b2}%
\path(1.40555,1.44126)(1.4105,1.43629)(1.41543,1.43131)
  (1.42034,1.4263)(1.42523,1.42127)(1.4301,1.41623)(1.43495,1.41116)
  (1.43978,1.40608)(1.44459,1.40098)(1.44938,1.39586)
  (1.45415,1.39072)(1.4589,1.38556)(1.46363,1.38038)(1.46834,1.37518)
  (1.47303,1.36997)(1.47769,1.36474)(1.48234,1.35949)
  (1.48697,1.35422)(1.49157,1.34893)(1.49616,1.34362)(1.50072,1.3383)
  (1.50527,1.33296)(1.50979,1.3276)(1.51429,1.32222)(1.51877,1.31683)
  (1.52323,1.31142)(1.52767,1.30599)(1.53209,1.30054)
  (1.53648,1.29508)(1.54086,1.28959)(1.54521,1.2841)(1.54954,1.27858)
  (1.55385,1.27305)(1.55814,1.2675)(1.5624,1.26193)(1.56664,1.25635)
  (1.57087,1.25075)(1.57506,1.24513)(1.57924,1.2395)(1.5834,1.23385)
  (1.58753,1.22819)(1.59164,1.22251)(1.59573,1.21681)(1.5998,1.2111)
  (1.60384,1.20537)(1.60786,1.19962)(1.61186,1.19386)
  (1.61583,1.18808)(1.61979,1.18229)(1.62372,1.17648)
  (1.62762,1.17066)(1.63151,1.16482)(1.63537,1.15897)(1.63921,1.1531)
  (1.64302,1.14722)(1.64682,1.14132)(1.65059,1.1354)(1.65433,1.12948)
  (1.65805,1.12353)(1.66175,1.11758)(1.66543,1.1116)(1.66908,1.10562)
  (1.67271,1.09962)(1.67631,1.0936)(1.6799,1.08757)(1.68345,1.08153)
  (1.68699,1.07547)(1.6905,1.0694)(1.69398,1.06332)(1.69745,1.05722)
  (1.70089,1.05111)(1.7043,1.04498)(1.70769,1.03885)(1.71106,1.03269)
  (1.7144,1.02653)(1.71772,1.02035)(1.72101,1.01416)(1.72428,1.00796)
  (1.72752,1.00174)(1.73074,0.995511)(1.73394,0.989269)
  (1.73711,0.983014)(1.74026,0.976748)(1.74338,0.970468)
  (1.74648,0.964177)(1.74955,0.957874)(1.7526,0.951558)
  (1.75562,0.945231)(1.75862,0.938892)(1.76159,0.932541)
  (1.76454,0.926178)(1.76746,0.919804)(1.77036,0.913418)
  (1.77324,0.907022)(1.77608,0.900614)(1.77891,0.894194)
  (1.7817,0.887764)(1.78448,0.881323)(1.78722,0.874871)
  (1.78995,0.868408)(1.79264,0.861935)(1.79531,0.855451)
  (1.79796,0.848956)(1.80058,0.842452)(1.80317,0.835937)
  (1.80574,0.829412)(1.80828,0.822876)(1.8108,0.816331)
  (1.81329,0.809777)(1.81576,0.803212)(1.8182,0.796638)
  (1.82061,0.790054)(1.823,0.783461)(1.82536,0.776858)
  (1.8277,0.770246)(1.83001,0.763625)(1.8323,0.756995)
  (1.83455,0.750357)(1.83679,0.743709)(1.83899,0.737052)(1.84117,0.730387)
\path(1.823,0.783461)(1.8323,0.756995)
\path(1.76339,0.900424)(1.73866,0.877771)
\path(1.79663,0.912096)(1.74717,0.866791)
\path(1.82964,0.923554)(1.75569,0.855812)
\path(1.82993,0.905048)(1.7642,0.844832)
\path(1.83023,0.886541)(1.77271,0.833853)
\path(1.83052,0.868035)(1.78122,0.822873)
\path(1.83082,0.849528)(1.78974,0.811893)
\path(1.83111,0.831022)(1.79825,0.800914)
\path(1.83141,0.812515)(1.80676,0.789934)
\path(1.8317,0.794008)(1.81527,0.778955)
\path(1.832,0.775502)(1.82378,0.767975)
\path(1.7303,0.888804)(1.83196,0.777832)
\path(1.74449,0.893787)(1.83163,0.798668)
\path(1.75868,0.89877)(1.8313,0.819504)
\path(1.77287,0.903753)(1.83096,0.840341)
\path(1.78706,0.908737)(1.83063,0.861177)
\path(1.80125,0.91372)(1.8303,0.882013)
\path(1.81544,0.918703)(1.82997,0.90285)
\path(1.77989,0.906218)(1.73015,0.888751)(1.8323,0.756995)
  (1.82963,0.923686)(1.77989,0.906218)
\color{rgb_b2b2ff}%
\path(0.142857,0.178571)(2.86016,1.06148)
\path(2.72023,0.98035)(2.73823,0.971466)
\path(2.71441,0.998234)(2.75042,0.980467)
\path(2.7086,1.01612)(2.76262,0.989468)
\path(2.70279,1.034)(2.77481,0.99847)
\path(2.69698,1.05189)(2.787,1.00747)
\path(2.7056,1.06265)(2.7992,1.01647)
\path(2.73651,1.06242)(2.81139,1.02547)
\path(2.76743,1.06218)(2.82358,1.03447)
\path(2.79834,1.06195)(2.83578,1.04347)
\path(2.82925,1.06171)(2.84797,1.05248)
\path(2.83623,1.04381)(2.84501,1.06159)
\path(2.81231,1.02615)(2.82985,1.06171)
\path(2.78838,1.00849)(2.8147,1.06182)
\path(2.76445,0.990826)(2.79954,1.06194)
\path(2.74053,0.973163)(2.78439,1.06205)
\path(2.72365,0.969796)(2.76923,1.06217)
\path(2.71761,0.988385)(2.75408,1.06228)
\path(2.71157,1.00697)(2.73892,1.0624)
\path(2.70553,1.02556)(2.72376,1.06251)
\path(2.69949,1.04415)(2.70861,1.06263)
\path(2.70975,1.0126)(2.72604,0.962465)(2.86016,1.06148)
  (2.69345,1.06274)(2.70975,1.0126)
\color{rgb_cc0000}%
\path(0.142857,0.178571)(3.71429,3.75)
\color{rgb_000000}%
\path(0.142857,0.178571)(3.71429,0.178571)
\path(3.70031,0.173913)(3.70111,0.182965)
\path(3.68634,0.169255)(3.68793,0.187358)
\path(3.67236,0.164597)(3.67475,0.191751)
\path(3.65839,0.159939)(3.66157,0.196144)
\path(3.64442,0.155281)(3.64839,0.200538)
\path(3.63044,0.150623)(3.63521,0.204931)
\path(3.61647,0.145965)(3.62203,0.209324)
\path(3.60249,0.141307)(3.60885,0.213717)
\path(3.58852,0.136649)(3.59567,0.218111)
\path(3.57454,0.131991)(3.58249,0.222504)
\path(3.56057,0.127333)(3.56931,0.226897)
\path(3.6098,0.213401)(3.55613,0.218111)
\path(3.66347,0.195511)(3.55613,0.204931)
\path(3.71262,0.178017)(3.55613,0.191751)
\path(3.68132,0.167584)(3.55613,0.178571)
\path(3.65003,0.157151)(3.55613,0.165392)
\path(3.61873,0.146718)(3.55613,0.152212)
\path(3.58743,0.136285)(3.55613,0.139032)
\path(3.55613,0.178571)(3.55613,0.125852)(3.71429,0.178571)
  (3.55613,0.23129)(3.55613,0.178571)
\path(0.142857,0.178571)(0.142857,3.75)
\path(0.14725,3.73682)(0.13752,3.73399)
\path(0.151644,3.72364)(0.132182,3.71798)
\path(0.156037,3.71046)(0.126845,3.70196)
\path(0.16043,3.69728)(0.121507,3.68595)
\path(0.164823,3.6841)(0.11617,3.66994)
\path(0.169217,3.67092)(0.110832,3.65393)
\path(0.17361,3.65774)(0.105495,3.63791)
\path(0.178003,3.64456)(0.100157,3.6219)
\path(0.182396,3.63138)(0.0948199,3.60589)
\path(0.18679,3.6182)(0.0962402,3.59184)
\path(0.191183,3.60502)(0.145908,3.59184)
\path(0.0971737,3.61295)(0.103318,3.59184)
\path(0.104209,3.63406)(0.116498,3.59184)
\path(0.111245,3.65516)(0.129677,3.59184)
\path(0.11828,3.67627)(0.142857,3.59184)
\path(0.125316,3.69738)(0.156037,3.59184)
\path(0.132351,3.71848)(0.169217,3.59184)
\path(0.139387,3.73959)(0.182396,3.59184)
\path(0.142857,3.59184)(0.195576,3.59184)(0.142857,3.75)
  (0.0901382,3.59184)(0.142857,3.59184)
\put(0.107711,0.178571){\makebox(0,0)[tr]{\hbox{\color{rgb_000000}\scriptsize $0$}}}
\put(3.78458,0.178571){\makebox(0,0)[l]{\hbox{\color{rgb_000000}\scriptsize $x_{1}$}}}
\put(0.142857,3.82029){\makebox(0,0)[b]{\hbox{\color{rgb_000000}\scriptsize $x_{2}$}}}
\put(3.64399,3.85544){\makebox(0,0)[tr]{\hbox{\color{rgb_000000}\scriptsize \rotatebox{45}{\textcolor{red!80!black}{$x_{1}=x_{2}$}}}}}
\put(1.11104,1.73629){\makebox(0,0)[bl]{\hbox{\color{rgb_000000}\scriptsize $\theta$}}}
\put(1.70057,1.14675){\makebox(0,0)[bl]{\hbox{\color{rgb_000000}\scriptsize \textcolor{black!30!white}{$-\theta$}}}}
\put(0.699598,2.21655){\makebox(0,0)[r]{\hbox{\color{rgb_000000}\scriptsize $r$}}}
\put(1.02576,2.96617){\makebox(0,0)[b]{\hbox{\color{rgb_000000}\scriptsize \textcolor{blue}{$(x_{1},x_{2})$}}}}
\put(2.93045,1.06148){\makebox(0,0)[l]{\hbox{\color{rgb_000000}\scriptsize \textcolor{blue!30!white}{$(x_{2},x_{1})$}}}}
\end{picture}%

%% file: figure3.eepic
%% Generated from figure3.xp on Fri Sep 16 13:10:45 JST 2022 by
%% ePiX-1.2.18
%% 
%%   Cartesian bounding box: [0,8] x [0,4]
%%   Actual size: 8 x 4cm
%%   Figure offset: left by 0cm, down by 0cm
%% 
%% usepackages epic,eepic,xcolor
%% 
\xdefinecolor{rgb_000000}{rgb}{0,0,0}%
\xdefinecolor{rgb_0000ff}{rgb}{0,0,1}%
\setlength{\unitlength}{1cm}%
\begin{picture}(8,4)(0,0)%
\path(0.307692,0.923077)(3.69231,0.923077)
\path(3.53415,0.883538)(3.57369,0.883538)
\path(3.53415,0.896717)(3.61323,0.896717)
\path(3.53415,0.909897)(3.65277,0.909897)
\path(3.53415,0.923077)(3.69231,0.923077)
\path(3.53415,0.936257)(3.65277,0.936257)
\path(3.53415,0.949436)(3.61323,0.949436)
\path(3.53415,0.962616)(3.57369,0.962616)
\path(3.67913,0.918684)(3.67913,0.92747)
\path(3.66595,0.91429)(3.66595,0.931863)
\path(3.65277,0.909897)(3.65277,0.936257)
\path(3.63959,0.905504)(3.63959,0.94065)
\path(3.62641,0.901111)(3.62641,0.945043)
\path(3.61323,0.896717)(3.61323,0.949436)
\path(3.60005,0.892324)(3.60005,0.95383)
\path(3.58687,0.887931)(3.58687,0.958223)
\path(3.57369,0.883538)(3.57369,0.962616)
\path(3.56051,0.879144)(3.56051,0.967009)
\path(3.54733,0.874751)(3.54733,0.971403)
\path(3.53415,0.870358)(3.53415,0.975796)
\path(3.53415,0.923077)(3.53415,0.870358)(3.69231,0.923077)
  (3.53415,0.975796)(3.53415,0.923077)
\path(2,0)(2,3.76923)
\path(2.04722,3.61107)(1.99341,3.74946)
\path(2.03294,3.61107)(1.98682,3.72969)
\path(2.01866,3.61107)(1.98023,3.70992)
\path(2.00439,3.61107)(1.97364,3.69015)
\path(1.99011,3.61107)(1.96705,3.67038)
\path(1.97583,3.61107)(1.96046,3.65061)
\path(1.96156,3.61107)(1.95387,3.63084)
\path(2.00439,3.75605)(1.9943,3.75213)
\path(2.00879,3.74287)(1.9886,3.73502)
\path(2.01318,3.72969)(1.9829,3.71792)
\path(2.01757,3.71651)(1.97719,3.70081)
\path(2.02197,3.70333)(1.97149,3.68371)
\path(2.02636,3.69015)(1.96579,3.6666)
\path(2.03075,3.67697)(1.96009,3.6495)
\path(2.03515,3.66379)(1.95439,3.63239)
\path(2.03954,3.65061)(1.94869,3.61529)
\path(2.04393,3.63743)(1.97614,3.61107)
\path(2.04833,3.62425)(2.01443,3.61107)
\path(2,3.61107)(2.05272,3.61107)(2,3.76923)(1.94728,3.61107)(2,3.61107)
\color{rgb_0000ff}%
\path(0.446012,0)(0.446355,0.00461538)(0.446701,0.00923077)
  (0.447049,0.0138462)(0.4474,0.0184615)(0.447754,0.0230769)
  (0.448111,0.0276923)(0.44847,0.0323077)(0.448832,0.0369231)
  (0.449196,0.0415385)(0.449564,0.0461538)(0.449934,0.0507692)
  (0.450308,0.0553846)(0.450684,0.06)(0.451063,0.0646154)
  (0.451446,0.0692308)(0.451831,0.0738462)(0.452219,0.0784615)
  (0.452611,0.0830769)(0.453006,0.0876923)(0.453404,0.0923077)
  (0.453805,0.0969231)(0.45421,0.101538)(0.454618,0.106154)
  (0.45503,0.110769)(0.455445,0.115385)(0.455863,0.12)
  (0.456286,0.124615)(0.456711,0.129231)(0.457141,0.133846)
  (0.457574,0.138462)(0.458011,0.143077)(0.458452,0.147692)
  (0.458896,0.152308)(0.459345,0.156923)(0.459797,0.161538)
  (0.460254,0.166154)(0.460715,0.170769)(0.46118,0.175385)
  (0.461649,0.18)(0.462123,0.184615)(0.462601,0.189231)
  (0.463084,0.193846)(0.463571,0.198462)(0.464062,0.203077)
  (0.464558,0.207692)(0.46506,0.212308)(0.465565,0.216923)
  (0.466076,0.221538)(0.466592,0.226154)(0.467113,0.230769)
  (0.467639,0.235385)(0.46817,0.24)(0.468706,0.244615)
  (0.469248,0.249231)(0.469796,0.253846)(0.470349,0.258462)
  (0.470908,0.263077)(0.471472,0.267692)(0.472042,0.272308)
  (0.472619,0.276923)(0.473201,0.281538)(0.47379,0.286154)
  (0.474385,0.290769)(0.474986,0.295385)(0.475594,0.3)
  (0.476209,0.304615)(0.476831,0.309231)(0.477459,0.313846)
  (0.478094,0.318462)(0.478737,0.323077)(0.479387,0.327692)
  (0.480044,0.332308)(0.480709,0.336923)(0.481381,0.341538)
  (0.482062,0.346154)(0.48275,0.350769)(0.483447,0.355385)
  (0.484152,0.36)(0.484866,0.364615)(0.485588,0.369231)
  (0.486319,0.373846)(0.48706,0.378462)(0.487809,0.383077)
  (0.488568,0.387692)(0.489337,0.392308)(0.490115,0.396923)
  (0.490903,0.401538)(0.491702,0.406154)(0.492511,0.410769)
  (0.493331,0.415385)(0.494162,0.42)(0.495005,0.424615)
  (0.495858,0.429231)(0.496723,0.433846)(0.497601,0.438462)
  (0.49849,0.443077)(0.499392,0.447692)(0.500307,0.452308)
  (0.501236,0.456923)(0.502177,0.461538)(0.503133,0.466154)
  (0.504102,0.470769)(0.505086,0.475385)(0.506085,0.48)
  (0.507099,0.484615)(0.508129,0.489231)(0.509174,0.493846)
  (0.510236,0.498462)(0.511315,0.503077)(0.512411,0.507692)
  (0.513525,0.512308)(0.514657,0.516923)(0.515808,0.521538)
  (0.516978,0.526154)(0.518168,0.530769)(0.519379,0.535385)
  (0.52061,0.54)(0.521862,0.544615)(0.523137,0.549231)
  (0.524435,0.553846)(0.525756,0.558462)(0.527101,0.563077)
  (0.528471,0.567692)(0.529866,0.572308)(0.531289,0.576923)
  (0.532738,0.581538)(0.534216,0.586154)(0.535722,0.590769)
  (0.537259,0.595385)(0.538827,0.6)(0.540426,0.604615)
  (0.542059,0.609231)(0.543726,0.613846)(0.545429,0.618462)
  (0.547168,0.623077)(0.548945,0.627692)(0.550762,0.632308)
  (0.552619,0.636923)(0.554519,0.641538)(0.556463,0.646154)
  (0.558452,0.650769)(0.560489,0.655385)(0.562575,0.66)
  (0.564712,0.664615)(0.566902,0.669231)(0.569147,0.673846)
  (0.571451,0.678462)(0.573814,0.683077)(0.576241,0.687692)
  (0.578732,0.692308)(0.581292,0.696923)(0.583924,0.701538)
  (0.58663,0.706154)(0.589415,0.710769)(0.592281,0.715385)
  (0.595233,0.72)(0.598275,0.724615)(0.601412,0.729231)
  (0.604648,0.733846)(0.607988,0.738462)(0.611438,0.743077)
  (0.615004,0.747692)(0.618692,0.752308)(0.622509,0.756923)
  (0.626462,0.761538)(0.630559,0.766154)(0.634808,0.770769)
  (0.639219,0.775385)(0.643802,0.78)(0.648567,0.784615)
  (0.653526,0.789231)(0.658692,0.793846)(0.664078,0.798462)
  (0.669701,0.803077)(0.675575,0.807692)(0.68172,0.812308)
  (0.688155,0.816923)(0.694902,0.821538)(0.701986,0.826154)
  (0.709432,0.830769)(0.717271,0.835385)(0.725535,0.84)
  (0.73426,0.844615)(0.743488,0.849231)(0.753264,0.853846)
  (0.763639,0.858462)(0.774669,0.863077)(0.786421,0.867692)
  (0.798968,0.872308)(0.812393,0.876923)(0.826791,0.881538)
  (0.84227,0.886154)(0.858956,0.890769)(0.876991,0.895385)
  (0.896542,0.9)(0.917801,0.904615)(0.94099,0.909231)
  (0.96637,0.913846)(0.994244,0.918462)
\path(1.03369,0.924308)(1.04266,0.925538)(1.05187,0.926769)
  (1.06135,0.928)(1.07109,0.92923)(1.08111,0.930461)
  (1.09143,0.931692)(1.10204,0.932922)(1.11296,0.934153)
  (1.12421,0.935384)(1.1358,0.936615)(1.14773,0.937845)
  (1.16003,0.939076)(1.17271,0.940307)(1.18577,0.941537)
  (1.19925,0.942768)(1.21314,0.943999)(1.22747,0.945229)
  (1.24226,0.94646)(1.25751,0.947691)(1.27325,0.948921)
  (1.28949,0.950152)(1.30626,0.951383)(1.32356,0.952614)
  (1.34142,0.953844)(1.35985,0.955075)(1.37888,0.956306)
  (1.39851,0.957536)(1.41877,0.958767)(1.43967,0.959998)
  (1.46122,0.961228)(1.48345,0.962459)(1.50636,0.96369)
  (1.52996,0.96492)(1.55427,0.966151)(1.57929,0.967382)
  (1.60502,0.968613)(1.63148,0.969843)(1.65864,0.971074)
  (1.68652,0.972305)(1.71509,0.973535)(1.74436,0.974766)
  (1.77429,0.975997)(1.80486,0.977227)(1.83605,0.978458)
  (1.86783,0.979689)(1.90015,0.980919)(1.93296,0.98215)
  (1.96623,0.983381)(1.99989,0.984612)(2.0339,0.985842)
  (2.06817,0.987073)(2.10266,0.988304)(2.1373,0.989534)
  (2.172,0.990765)(2.20671,0.991996)(2.24136,0.993226)
  (2.27587,0.994457)(2.31017,0.995688)(2.3442,0.996918)
  (2.3779,0.998149)(2.41122,0.99938)(2.44408,1.00061)
  (2.47646,1.00184)(2.50829,1.00307)(2.53955,1.0043)(2.5702,1.00553)
  (2.6002,1.00676)(2.62954,1.00799)(2.6582,1.00923)(2.68616,1.01046)
  (2.71342,1.01169)(2.73996,1.01292)(2.76579,1.01415)(2.7909,1.01538)
  (2.81531,1.01661)(2.83901,1.01784)(2.86202,1.01907)(2.88435,1.0203)
  (2.90601,1.02153)(2.92701,1.02276)(2.94738,1.02399)
  (2.96712,1.02522)(2.98625,1.02646)(3.00479,1.02769)
  (3.02276,1.02892)(3.04018,1.03015)(3.05706,1.03138)
  (3.07341,1.03261)(3.08927,1.03384)(3.10464,1.03507)(3.11954,1.0363)
  (3.13399,1.03753)(3.14801,1.03876)(3.1616,1.03999)(3.17479,1.04122)
  (3.18759,1.04245)(3.20001,1.04368)(3.21208,1.04492)
  (3.22379,1.04615)(3.23517,1.04738)(3.24623,1.04861)
  (3.25697,1.04984)(3.26742,1.05107)(3.27758,1.0523)(3.28746,1.05353)
  (3.29708,1.05476)(3.30644,1.05599)(3.31555,1.05722)
  (3.32442,1.05845)(3.33306,1.05968)(3.34148,1.06091)
  (3.34969,1.06215)(3.35769,1.06338)(3.36549,1.06461)(3.3731,1.06584)
  (3.38053,1.06707)(3.38778,1.0683)(3.39486,1.06953)(3.40177,1.07076)
  (3.40852,1.07199)(3.41512,1.07322)(3.42156,1.07445)
  (3.42787,1.07568)(3.43403,1.07691)(3.44006,1.07814)
  (3.44596,1.07937)(3.45173,1.08061)(3.45738,1.08184)
  (3.46292,1.08307)(3.46834,1.0843)(3.47365,1.08553)(3.47885,1.08676)
  (3.48395,1.08799)(3.48895,1.08922)(3.49385,1.09045)
  (3.49866,1.09168)(3.50338,1.09291)(3.50801,1.09414)
  (3.51255,1.09537)(3.51701,1.0966)(3.52139,1.09784)(3.5257,1.09907)
  (3.52992,1.1003)(3.53408,1.10153)(3.53816,1.10276)(3.54217,1.10399)
  (3.54612,1.10522)(3.55,1.10645)(3.55382,1.10768)(3.55758,1.10891)
  (3.56127,1.11014)(3.56491,1.11137)(3.56849,1.1126)(3.57202,1.11383)
  (3.5755,1.11506)(3.57892,1.1163)(3.58229,1.11753)(3.58561,1.11876)
  (3.58889,1.11999)(3.59212,1.12122)(3.5953,1.12245)(3.59844,1.12368)
  (3.60154,1.12491)(3.60459,1.12614)(3.60761,1.12737)(3.61059,1.1286)
  (3.61352,1.12983)(3.61643,1.13106)(3.61929,1.13229)
  (3.62212,1.13353)(3.62491,1.13476)(3.62767,1.13599)(3.6304,1.13722)
  (3.6331,1.13845)(3.63577,1.13968)(3.6384,1.14091)(3.64101,1.14214)
  (3.64359,1.14337)(3.64614,1.1446)(3.64866,1.14583)(3.65116,1.14706)
  (3.65363,1.14829)(3.65607,1.14952)(3.65849,1.15075)
  (3.66089,1.15199)(3.66327,1.15322)(3.66562,1.15445)
  (3.66795,1.15568)(3.67026,1.15691)(3.67254,1.15814)
  (3.67481,1.15937)(3.67706,1.1606)(3.67929,1.16183)(3.6815,1.16306)
  (3.68369,1.16429)(3.68586,1.16552)(3.68802,1.16675)
  (3.69016,1.16798)(3.69228,1.16922)
\path(0.307719,1.16925)(0.313995,1.17294)(0.320139,1.17663)
  (0.326164,1.18032)(0.332078,1.18401)(0.337893,1.18771)
  (0.343616,1.1914)(0.349258,1.19509)(0.354825,1.19878)
  (0.360328,1.20248)(0.365773,1.20617)(0.371168,1.20986)
  (0.37652,1.21355)(0.381837,1.21724)(0.387125,1.22094)
  (0.392391,1.22463)(0.397643,1.22832)(0.402887,1.23201)
  (0.408129,1.2357)(0.413376,1.2394)(0.418635,1.24309)
  (0.423912,1.24678)(0.429215,1.25047)(0.43455,1.25417)
  (0.439924,1.25786)(0.445344,1.26155)(0.450818,1.26524)
  (0.456354,1.26893)(0.461958,1.27263)(0.46764,1.27632)
  (0.473408,1.28001)(0.47927,1.2837)(0.485236,1.2874)
  (0.491316,1.29109)(0.49752,1.29478)(0.503859,1.29847)
  (0.510345,1.30216)(0.516989,1.30586)(0.523806,1.30955)
  (0.53081,1.31324)(0.538014,1.31693)(0.545437,1.32062)
  (0.553095,1.32432)(0.561007,1.32801)(0.569194,1.3317)
  (0.577678,1.33539)(0.586484,1.33909)(0.595637,1.34278)
  (0.605168,1.34647)(0.615107,1.35016)(0.62549,1.35385)
  (0.636354,1.35755)(0.647743,1.36124)(0.659703,1.36493)
  (0.672285,1.36862)(0.685548,1.37231)(0.699556,1.37601)
  (0.714379,1.3797)(0.730099,1.38339)(0.746805,1.38708)
  (0.764597,1.39078)(0.78359,1.39447)(0.80391,1.39816)
  (0.825702,1.40185)(0.849128,1.40554)(0.874373,1.40924)
  (0.901644,1.41293)(0.931175,1.41662)(0.963231,1.42031)
  (0.998108,1.424)(1.03614,1.4277)(1.07768,1.43139)(1.12313,1.43508)
  (1.17292,1.43877)(1.22747,1.44247)(1.28723,1.44616)
  (1.35258,1.44985)(1.42384,1.45354)(1.50117,1.45723)
  (1.58453,1.46093)(1.67358,1.46462)(1.76765,1.46831)(1.86567,1.472)
  (1.96626,1.47569)(2.06775,1.47939)(2.16842,1.48308)(2.2666,1.48677)
  (2.36089,1.49046)(2.4502,1.49416)(2.53383,1.49785)(2.61142,1.50154)
  (2.68292,1.50523)(2.74848,1.50892)(2.8084,1.51262)(2.86309,1.51631)
  (2.91296,1.52)(2.95845,1.52369)(3,1.52739)(3.03798,1.53108)
  (3.07278,1.53477)(3.10472,1.53846)(3.1341,1.54215)(3.16118,1.54585)
  (3.18621,1.54954)(3.20939,1.55323)(3.2309,1.55692)(3.25091,1.56061)
  (3.26957,1.56431)(3.287,1.568)(3.30331,1.57169)(3.31862,1.57538)
  (3.33299,1.57908)(3.34653,1.58277)(3.3593,1.58646)(3.37135,1.59015)
  (3.38276,1.59384)(3.39357,1.59754)(3.40383,1.60123)
  (3.41358,1.60492)(3.42286,1.60861)(3.4317,1.6123)(3.44014,1.616)
  (3.4482,1.61969)(3.4559,1.62338)(3.46328,1.62707)(3.47035,1.63077)
  (3.47714,1.63446)(3.48366,1.63815)(3.48993,1.64184)
  (3.49596,1.64553)(3.50176,1.64923)(3.50736,1.65292)
  (3.51277,1.65661)(3.51798,1.6603)(3.52303,1.66399)(3.5279,1.66769)
  (3.53262,1.67138)(3.53719,1.67507)(3.54163,1.67876)
  (3.54592,1.68246)(3.5501,1.68615)(3.55415,1.68984)(3.55808,1.69353)
  (3.56191,1.69722)(3.56564,1.70092)(3.56927,1.70461)(3.5728,1.7083)
  (3.57624,1.71199)(3.5796,1.71568)(3.58288,1.71938)(3.58608,1.72307)
  (3.5892,1.72676)(3.59226,1.73045)(3.59524,1.73415)(3.59816,1.73784)
  (3.60102,1.74153)(3.60382,1.74522)(3.60656,1.74891)
  (3.60924,1.75261)(3.61187,1.7563)(3.61445,1.75999)(3.61699,1.76368)
  (3.61947,1.76738)(3.62192,1.77107)(3.62431,1.77476)
  (3.62667,1.77845)(3.62899,1.78214)(3.63127,1.78584)
  (3.63351,1.78953)(3.63572,1.79322)(3.63789,1.79691)(3.64003,1.8006)
  (3.64214,1.8043)(3.64422,1.80799)(3.64627,1.81168)(3.64829,1.81537)
  (3.65029,1.81907)(3.65225,1.82276)(3.6542,1.82645)(3.65612,1.83014)
  (3.65802,1.83383)(3.65989,1.83753)(3.66174,1.84122)
  (3.66358,1.84491)(3.66539,1.8486)(3.66718,1.85229)(3.66896,1.85599)
  (3.67072,1.85968)(3.67246,1.86337)(3.67419,1.86706)(3.6759,1.87076)
  (3.6776,1.87445)(3.67928,1.87814)(3.68095,1.88183)(3.6826,1.88552)
  (3.68425,1.88922)(3.68588,1.89291)(3.6875,1.8966)(3.68911,1.90029)
  (3.69071,1.90398)(3.6923,1.90768)
\path(0.307699,1.90771)(0.310331,1.91386)(0.312941,1.92002)
  (0.31553,1.92617)(0.318102,1.93232)(0.320658,1.93848)
  (0.3232,1.94463)(0.325731,1.95078)(0.328253,1.95694)
  (0.330768,1.96309)(0.333277,1.96924)(0.335783,1.9754)
  (0.338288,1.98155)(0.340794,1.98771)(0.343303,1.99386)
  (0.345817,2.00001)(0.348339,2.00617)(0.35087,2.01232)
  (0.353412,2.01847)(0.355968,2.02463)(0.35854,2.03078)
  (0.36113,2.03694)(0.363741,2.04309)(0.366375,2.04924)
  (0.369035,2.0554)(0.371723,2.06155)(0.374442,2.0677)
  (0.377196,2.07386)(0.379985,2.08001)(0.382815,2.08616)
  (0.385688,2.09232)(0.388608,2.09847)(0.391578,2.10463)
  (0.394602,2.11078)(0.397685,2.11693)(0.400829,2.12309)
  (0.404041,2.12924)(0.407324,2.13539)(0.410684,2.14155)
  (0.414127,2.1477)(0.417658,2.15386)(0.421284,2.16001)
  (0.425011,2.16616)(0.428847,2.17232)(0.4328,2.17847)
  (0.436878,2.18462)(0.441091,2.19078)(0.445448,2.19693)
  (0.449961,2.20308)(0.454641,2.20924)(0.459502,2.21539)
  (0.464558,2.22155)(0.469825,2.2277)(0.47532,2.23385)
  (0.481062,2.24001)(0.487073,2.24616)(0.493375,2.25231)
  (0.499996,2.25847)(0.506963,2.26462)(0.514311,2.27078)
  (0.522076,2.27693)(0.530299,2.28308)(0.539026,2.28924)
  (0.548312,2.29539)(0.558216,2.30154)(0.568809,2.3077)
  (0.580168,2.31385)(0.592387,2.32001)(0.605572,2.32616)
  (0.619847,2.33231)(0.635357,2.33847)(0.652272,2.34462)
  (0.670797,2.35077)(0.691171,2.35693)(0.713682,2.36308)
  (0.738679,2.36923)(0.766581,2.37539)(0.797902,2.38154)
  (0.833265,2.3877)(0.87344,2.39385)(0.919369,2.4)(0.972204,2.40616)
  (1.03334,2.41231)(1.10445,2.41846)(1.18744,2.42462)
  (1.28432,2.43077)(1.39698,2.43693)(1.52655,2.44308)
  (1.67262,2.44923)(1.83234,2.45539)(2.00004,2.46154)
  (2.16807,2.46769)(2.32869,2.47385)(2.47601,2.48)(2.60696,2.48615)
  (2.72094,2.49231)(2.81903,2.49846)(2.90303,2.50462)
  (2.97499,2.51077)(3.03683,2.51692)(3.09023,2.52308)
  (3.13661,2.52923)(3.17714,2.53538)(3.21279,2.54154)
  (3.24433,2.54769)(3.27239,2.55385)(3.29751,2.56)(3.3201,2.56615)
  (3.34052,2.57231)(3.35906,2.57846)(3.37597,2.58461)
  (3.39145,2.59077)(3.40568,2.59692)(3.4188,2.60307)(3.43093,2.60923)
  (3.44219,2.61538)(3.45267,2.62154)(3.46244,2.62769)
  (3.47159,2.63384)(3.48016,2.64)(3.48822,2.64615)(3.4958,2.6523)
  (3.50296,2.65846)(3.50973,2.66461)(3.51614,2.67077)
  (3.52222,2.67692)(3.52799,2.68307)(3.53349,2.68923)
  (3.53873,2.69538)(3.54373,2.70153)(3.54851,2.70769)
  (3.55309,2.71384)(3.55747,2.72)(3.56167,2.72615)(3.56571,2.7323)
  (3.5696,2.73846)(3.57333,2.74461)(3.57693,2.75076)(3.5804,2.75692)
  (3.58375,2.76307)(3.58699,2.76922)(3.59012,2.77538)
  (3.59315,2.78153)(3.59609,2.78769)(3.59893,2.79384)
  (3.60169,2.79999)(3.60437,2.80615)(3.60698,2.8123)(3.60951,2.81845)
  (3.61197,2.82461)(3.61437,2.83076)(3.61671,2.83692)
  (3.61899,2.84307)(3.62121,2.84922)(3.62338,2.85538)
  (3.62549,2.86153)(3.62756,2.86768)(3.62959,2.87384)
  (3.63157,2.87999)(3.63351,2.88614)(3.6354,2.8923)(3.63726,2.89845)
  (3.63909,2.90461)(3.64088,2.91076)(3.64263,2.91691)
  (3.64436,2.92307)(3.64605,2.92922)(3.64771,2.93537)
  (3.64935,2.94153)(3.65096,2.94768)(3.65254,2.95384)(3.6541,2.95999)
  (3.65563,2.96614)(3.65715,2.9723)(3.65864,2.97845)(3.66011,2.9846)
  (3.66156,2.99076)(3.66299,2.99691)(3.66441,3.00306)(3.6658,3.00922)
  (3.66718,3.01537)(3.66855,3.02153)(3.6699,3.02768)(3.67123,3.03383)
  (3.67256,3.03999)(3.67387,3.04614)(3.67516,3.05229)
  (3.67645,3.05845)(3.67772,3.0646)(3.67898,3.07076)(3.68023,3.07691)
  (3.68147,3.08306)(3.68271,3.08922)(3.68393,3.09537)
  (3.68515,3.10152)(3.68636,3.10768)(3.68756,3.11383)
  (3.68876,3.11999)(3.68994,3.12614)(3.69113,3.13229)(3.6923,3.13845)
\path(0.307695,3.13848)(0.308297,3.14163)(0.308897,3.14478)
  (0.309497,3.14794)(0.310095,3.15109)(0.310692,3.15425)
  (0.311288,3.1574)(0.311883,3.16055)(0.312477,3.16371)
  (0.313071,3.16686)(0.313663,3.17001)(0.314255,3.17317)
  (0.314846,3.17632)(0.315437,3.17948)(0.316027,3.18263)
  (0.316617,3.18578)(0.317206,3.18894)(0.317795,3.19209)
  (0.318383,3.19524)(0.318971,3.1984)(0.319559,3.20155)
  (0.320147,3.20471)(0.320735,3.20786)(0.321322,3.21101)
  (0.32191,3.21417)(0.322498,3.21732)(0.323085,3.22047)
  (0.323673,3.22363)(0.324261,3.22678)(0.32485,3.22994)
  (0.325438,3.23309)(0.326027,3.23624)(0.326617,3.2394)
  (0.327207,3.24255)(0.327797,3.24571)(0.328389,3.24886)
  (0.32898,3.25201)(0.329573,3.25517)(0.330166,3.25832)
  (0.33076,3.26147)(0.331355,3.26463)(0.331951,3.26778)
  (0.332548,3.27094)(0.333146,3.27409)(0.333745,3.27724)
  (0.334346,3.2804)(0.334947,3.28355)(0.33555,3.2867)
  (0.336154,3.28986)(0.33676,3.29301)(0.337367,3.29617)
  (0.337976,3.29932)(0.338586,3.30247)(0.339198,3.30563)
  (0.339812,3.30878)(0.340427,3.31193)(0.341045,3.31509)
  (0.341664,3.31824)(0.342286,3.3214)(0.342909,3.32455)
  (0.343535,3.3277)(0.344163,3.33086)(0.344793,3.33401)
  (0.345425,3.33716)(0.34606,3.34032)(0.346698,3.34347)
  (0.347338,3.34663)(0.347981,3.34978)(0.348626,3.35293)
  (0.349274,3.35609)(0.349926,3.35924)(0.35058,3.36239)
  (0.351237,3.36555)(0.351898,3.3687)(0.352562,3.37186)
  (0.353229,3.37501)(0.353899,3.37816)(0.354573,3.38132)
  (0.355251,3.38447)(0.355932,3.38762)(0.356618,3.39078)
  (0.357307,3.39393)(0.358,3.39709)(0.358697,3.40024)
  (0.359398,3.40339)(0.360104,3.40655)(0.360814,3.4097)
  (0.361529,3.41285)(0.362248,3.41601)(0.362972,3.41916)
  (0.363701,3.42232)(0.364435,3.42547)(0.365174,3.42862)
  (0.365918,3.43178)(0.366667,3.43493)(0.367422,3.43808)
  (0.368183,3.44124)(0.368949,3.44439)(0.369722,3.44755)
  (0.3705,3.4507)(0.371285,3.45385)(0.372075,3.45701)
  (0.372873,3.46016)(0.373677,3.46332)(0.374488,3.46647)
  (0.375305,3.46962)(0.37613,3.47278)(0.376962,3.47593)
  (0.377802,3.47908)(0.378649,3.48224)(0.379505,3.48539)
  (0.380368,3.48855)(0.381239,3.4917)(0.382119,3.49485)
  (0.383007,3.49801)(0.383905,3.50116)(0.384811,3.50431)
  (0.385727,3.50747)(0.386652,3.51062)(0.387587,3.51378)
  (0.388532,3.51693)(0.389487,3.52008)(0.390452,3.52324)
  (0.391429,3.52639)(0.392416,3.52954)(0.393415,3.5327)
  (0.394425,3.53585)(0.395447,3.53901)(0.396481,3.54216)
  (0.397528,3.54531)(0.398588,3.54847)(0.39966,3.55162)
  (0.400747,3.55477)(0.401847,3.55793)(0.402961,3.56108)
  (0.40409,3.56424)(0.405234,3.56739)(0.406394,3.57054)
  (0.407569,3.5737)(0.40876,3.57685)(0.409968,3.58)(0.411194,3.58316)
  (0.412437,3.58631)(0.413698,3.58947)(0.414977,3.59262)
  (0.416276,3.59577)(0.417594,3.59893)(0.418933,3.60208)
  (0.420293,3.60523)(0.421674,3.60839)(0.423077,3.61154)
  (0.424503,3.6147)(0.425953,3.61785)(0.427426,3.621)
  (0.428925,3.62416)(0.430449,3.62731)(0.431999,3.63046)
  (0.433577,3.63362)(0.435183,3.63677)(0.436818,3.63993)
  (0.438483,3.64308)(0.440179,3.64623)(0.441906,3.64939)
  (0.443667,3.65254)(0.445462,3.6557)(0.447292,3.65885)
  (0.449158,3.662)(0.451063,3.66516)(0.453006,3.66831)
  (0.454989,3.67146)(0.457014,3.67462)(0.459083,3.67777)
  (0.461196,3.68093)(0.463356,3.68408)(0.465564,3.68723)
  (0.467822,3.69039)(0.470132,3.69354)(0.472495,3.69669)
  (0.474915,3.69985)(0.477393,3.703)(0.479931,3.70616)
  (0.482532,3.70931)(0.485199,3.71246)(0.487933,3.71562)
  (0.490739,3.71877)(0.493618,3.72192)(0.496575,3.72508)
  (0.499612,3.72823)(0.502732,3.73139)(0.505941,3.73454)
  (0.509241,3.73769)(0.512637,3.74085)(0.516134,3.744)
  (0.519735,3.74715)(0.523446,3.75031)(0.527273,3.75346)
  (0.531221,3.75662)(0.535296,3.75977)(0.539504,3.76292)
  (0.543853,3.76608)(0.548349,3.76923)
\color{rgb_000000}%
\path(0.307692,1.16923)(0.341538,1.16923)
\path(0.409231,1.16923)(0.443077,1.16923)
\path(0.443077,1.16923)(0.476923,1.16923)
\path(0.544615,1.16923)(0.578462,1.16923)
\path(0.578462,1.16923)(0.612308,1.16923)
\path(0.68,1.16923)(0.713846,1.16923)
\path(0.713846,1.16923)(0.747692,1.16923)
\path(0.815385,1.16923)(0.849231,1.16923)
\path(0.849231,1.16923)(0.883077,1.16923)
\path(0.950769,1.16923)(0.984615,1.16923)
\path(0.984615,1.16923)(1.01846,1.16923)
\path(1.08615,1.16923)(1.12,1.16923)
\path(1.12,1.16923)(1.15385,1.16923)
\path(1.22154,1.16923)(1.25538,1.16923)
\path(1.25538,1.16923)(1.28923,1.16923)
\path(1.35692,1.16923)(1.39077,1.16923)
\path(1.39077,1.16923)(1.42462,1.16923)
\path(1.49231,1.16923)(1.52615,1.16923)
\path(1.52615,1.16923)(1.56,1.16923)
\path(1.62769,1.16923)(1.66154,1.16923)
\path(1.66154,1.16923)(1.69538,1.16923)
\path(1.76308,1.16923)(1.79692,1.16923)
\path(1.79692,1.16923)(1.83077,1.16923)
\path(1.89846,1.16923)(1.93231,1.16923)
\path(1.93231,1.16923)(1.96615,1.16923)
\path(2.03385,1.16923)(2.06769,1.16923)
\path(2.06769,1.16923)(2.10154,1.16923)
\path(2.16923,1.16923)(2.20308,1.16923)
\path(2.20308,1.16923)(2.23692,1.16923)
\path(2.30462,1.16923)(2.33846,1.16923)
\path(2.33846,1.16923)(2.37231,1.16923)
\path(2.44,1.16923)(2.47385,1.16923)
\path(2.47385,1.16923)(2.50769,1.16923)
\path(2.57538,1.16923)(2.60923,1.16923)
\path(2.60923,1.16923)(2.64308,1.16923)
\path(2.71077,1.16923)(2.74462,1.16923)
\path(2.74462,1.16923)(2.77846,1.16923)
\path(2.84615,1.16923)(2.88,1.16923)
\path(2.88,1.16923)(2.91385,1.16923)
\path(2.98154,1.16923)(3.01538,1.16923)
\path(3.01538,1.16923)(3.04923,1.16923)
\path(3.11692,1.16923)(3.15077,1.16923)
\path(3.15077,1.16923)(3.18462,1.16923)
\path(3.25231,1.16923)(3.28615,1.16923)
\path(3.28615,1.16923)(3.32,1.16923)
\path(3.38769,1.16923)(3.42154,1.16923)
\path(3.42154,1.16923)(3.45538,1.16923)
\path(3.52308,1.16923)(3.55692,1.16923)
\path(3.55692,1.16923)(3.59077,1.16923)
\path(3.65846,1.16923)(3.69231,1.16923)
\path(0.307692,1.90769)(0.341538,1.90769)
\path(0.409231,1.90769)(0.443077,1.90769)
\path(0.443077,1.90769)(0.476923,1.90769)
\path(0.544615,1.90769)(0.578462,1.90769)
\path(0.578462,1.90769)(0.612308,1.90769)
\path(0.68,1.90769)(0.713846,1.90769)
\path(0.713846,1.90769)(0.747692,1.90769)
\path(0.815385,1.90769)(0.849231,1.90769)
\path(0.849231,1.90769)(0.883077,1.90769)
\path(0.950769,1.90769)(0.984615,1.90769)
\path(0.984615,1.90769)(1.01846,1.90769)
\path(1.08615,1.90769)(1.12,1.90769)
\path(1.12,1.90769)(1.15385,1.90769)
\path(1.22154,1.90769)(1.25538,1.90769)
\path(1.25538,1.90769)(1.28923,1.90769)
\path(1.35692,1.90769)(1.39077,1.90769)
\path(1.39077,1.90769)(1.42462,1.90769)
\path(1.49231,1.90769)(1.52615,1.90769)
\path(1.52615,1.90769)(1.56,1.90769)
\path(1.62769,1.90769)(1.66154,1.90769)
\path(1.66154,1.90769)(1.69538,1.90769)
\path(1.76308,1.90769)(1.79692,1.90769)
\path(1.79692,1.90769)(1.83077,1.90769)
\path(1.89846,1.90769)(1.93231,1.90769)
\path(1.93231,1.90769)(1.96615,1.90769)
\path(2.03385,1.90769)(2.06769,1.90769)
\path(2.06769,1.90769)(2.10154,1.90769)
\path(2.16923,1.90769)(2.20308,1.90769)
\path(2.20308,1.90769)(2.23692,1.90769)
\path(2.30462,1.90769)(2.33846,1.90769)
\path(2.33846,1.90769)(2.37231,1.90769)
\path(2.44,1.90769)(2.47385,1.90769)
\path(2.47385,1.90769)(2.50769,1.90769)
\path(2.57538,1.90769)(2.60923,1.90769)
\path(2.60923,1.90769)(2.64308,1.90769)
\path(2.71077,1.90769)(2.74462,1.90769)
\path(2.74462,1.90769)(2.77846,1.90769)
\path(2.84615,1.90769)(2.88,1.90769)
\path(2.88,1.90769)(2.91385,1.90769)
\path(2.98154,1.90769)(3.01538,1.90769)
\path(3.01538,1.90769)(3.04923,1.90769)
\path(3.11692,1.90769)(3.15077,1.90769)
\path(3.15077,1.90769)(3.18462,1.90769)
\path(3.25231,1.90769)(3.28615,1.90769)
\path(3.28615,1.90769)(3.32,1.90769)
\path(3.38769,1.90769)(3.42154,1.90769)
\path(3.42154,1.90769)(3.45538,1.90769)
\path(3.52308,1.90769)(3.55692,1.90769)
\path(3.55692,1.90769)(3.59077,1.90769)
\path(3.65846,1.90769)(3.69231,1.90769)
\path(0.307692,3.13846)(0.341538,3.13846)
\path(0.409231,3.13846)(0.443077,3.13846)
\path(0.443077,3.13846)(0.476923,3.13846)
\path(0.544615,3.13846)(0.578462,3.13846)
\path(0.578462,3.13846)(0.612308,3.13846)
\path(0.68,3.13846)(0.713846,3.13846)
\path(0.713846,3.13846)(0.747692,3.13846)
\path(0.815385,3.13846)(0.849231,3.13846)
\path(0.849231,3.13846)(0.883077,3.13846)
\path(0.950769,3.13846)(0.984615,3.13846)
\path(0.984615,3.13846)(1.01846,3.13846)
\path(1.08615,3.13846)(1.12,3.13846)
\path(1.12,3.13846)(1.15385,3.13846)
\path(1.22154,3.13846)(1.25538,3.13846)
\path(1.25538,3.13846)(1.28923,3.13846)
\path(1.35692,3.13846)(1.39077,3.13846)
\path(1.39077,3.13846)(1.42462,3.13846)
\path(1.49231,3.13846)(1.52615,3.13846)
\path(1.52615,3.13846)(1.56,3.13846)
\path(1.62769,3.13846)(1.66154,3.13846)
\path(1.66154,3.13846)(1.69538,3.13846)
\path(1.76308,3.13846)(1.79692,3.13846)
\path(1.79692,3.13846)(1.83077,3.13846)
\path(1.89846,3.13846)(1.93231,3.13846)
\path(1.93231,3.13846)(1.96615,3.13846)
\path(2.03385,3.13846)(2.06769,3.13846)
\path(2.06769,3.13846)(2.10154,3.13846)
\path(2.16923,3.13846)(2.20308,3.13846)
\path(2.20308,3.13846)(2.23692,3.13846)
\path(2.30462,3.13846)(2.33846,3.13846)
\path(2.33846,3.13846)(2.37231,3.13846)
\path(2.44,3.13846)(2.47385,3.13846)
\path(2.47385,3.13846)(2.50769,3.13846)
\path(2.57538,3.13846)(2.60923,3.13846)
\path(2.60923,3.13846)(2.64308,3.13846)
\path(2.71077,3.13846)(2.74462,3.13846)
\path(2.74462,3.13846)(2.77846,3.13846)
\path(2.84615,3.13846)(2.88,3.13846)
\path(2.88,3.13846)(2.91385,3.13846)
\path(2.98154,3.13846)(3.01538,3.13846)
\path(3.01538,3.13846)(3.04923,3.13846)
\path(3.11692,3.13846)(3.15077,3.13846)
\path(3.15077,3.13846)(3.18462,3.13846)
\path(3.25231,3.13846)(3.28615,3.13846)
\path(3.28615,3.13846)(3.32,3.13846)
\path(3.38769,3.13846)(3.42154,3.13846)
\path(3.42154,3.13846)(3.45538,3.13846)
\path(3.52308,3.13846)(3.55692,3.13846)
\path(3.55692,3.13846)(3.59077,3.13846)
\path(3.65846,3.13846)(3.69231,3.13846)
\put(1.96485,0.872546){\makebox(0,0)[tr]{\hbox{\color{rgb_000000}\scriptsize $0$}}}
\put(3.7626,0.923077){\makebox(0,0)[l]{\hbox{\color{rgb_000000}\scriptsize $g_{0}$}}}
\put(2,3.83952){\makebox(0,0)[b]{\hbox{\color{rgb_000000}\scriptsize $\lambda$}}}
\put(1.02497,0.852785){\makebox(0,0)[t]{\hbox{\color{rgb_000000}\scriptsize $g_{\ast}$}}}
\put(2.03545,1.23952){\makebox(0,0)[bl]{\hbox{\color{rgb_000000}\scriptsize $4^{2}$}}}
\put(2.03545,1.97798){\makebox(0,0)[bl]{\hbox{\color{rgb_000000}\scriptsize $8^{2}$}}}
\put(2.03545,3.20875){\makebox(0,0)[bl]{\hbox{\color{rgb_000000}\scriptsize $12^{2}$}}}
\put(3.72743,1.20436){\makebox(0,0)[b]{\hbox{\color{rgb_000000}\scriptsize \textcolor{blue}{$\lambda_{0}$}}}}
\put(3.72745,1.94282){\makebox(0,0)[b]{\hbox{\color{rgb_000000}\scriptsize \textcolor{blue}{$\lambda_{1}$}}}}
\put(3.72745,3.17359){\makebox(0,0)[b]{\hbox{\color{rgb_000000}\scriptsize \textcolor{blue}{$\lambda_{2}$}}}}
\put(0.2374,0.817639){\makebox(0,0)[t]{\hbox{\color{rgb_000000}\scriptsize $-\infty$}}}
\put(3.69231,0.817639){\makebox(0,0)[t]{\hbox{\color{rgb_000000}\scriptsize $+\infty$}}}
\path(4.30769,0.461538)(7.69231,0.461538)
\path(7.54853,0.509465)(7.53415,0.496293)
\path(7.56291,0.504672)(7.53415,0.478329)
\path(7.57728,0.49988)(7.53415,0.460366)
\path(7.59166,0.495087)(7.53415,0.442402)
\path(7.60604,0.490294)(7.53415,0.424438)
\path(7.62042,0.485502)(7.53818,0.410161)
\path(7.6348,0.480709)(7.569,0.420437)
\path(7.64917,0.475916)(7.59983,0.430712)
\path(7.66355,0.471124)(7.63066,0.440988)
\path(7.67793,0.466331)(7.66148,0.451263)
\path(7.53415,0.429307)(7.54853,0.413612)
\path(7.53415,0.449795)(7.56291,0.418405)
\path(7.53415,0.470282)(7.57728,0.423197)
\path(7.53415,0.49077)(7.59166,0.42799)
\path(7.53415,0.511258)(7.60604,0.432783)
\path(7.55721,0.50657)(7.62042,0.437575)
\path(7.58423,0.497564)(7.6348,0.442368)
\path(7.61125,0.488557)(7.64917,0.447161)
\path(7.63827,0.479551)(7.66355,0.451953)
\path(7.66529,0.470545)(7.67793,0.456746)
\path(7.53415,0.461538)(7.53415,0.408819)(7.69231,0.461538)
  (7.53415,0.514257)(7.53415,0.461538)
\path(6,0)(6,3.76923)
\path(5.95167,3.62425)(5.97839,3.61107)
\path(5.95607,3.63743)(6.00949,3.61107)
\path(5.96046,3.65061)(6.0406,3.61107)
\path(5.96485,3.66379)(6.04898,3.62228)
\path(5.96925,3.67697)(6.04286,3.64065)
\path(5.97364,3.69015)(6.03674,3.65902)
\path(5.97803,3.70333)(6.03061,3.67739)
\path(5.98243,3.71651)(6.02449,3.69576)
\path(5.98682,3.72969)(6.01837,3.71413)
\path(5.99121,3.74287)(6.01225,3.73249)
\path(5.99561,3.75605)(6.00612,3.75086)
\path(6.03819,3.61107)(6.04686,3.62865)
\path(6.02366,3.61107)(6.041,3.64622)
\path(6.00913,3.61107)(6.03515,3.66379)
\path(5.99461,3.61107)(6.02929,3.68137)
\path(5.98008,3.61107)(6.02343,3.69894)
\path(5.96555,3.61107)(6.01757,3.71651)
\path(5.95102,3.61107)(6.01172,3.73408)
\path(5.96975,3.67847)(6.00586,3.75166)
\path(6,3.61107)(6.05272,3.61107)(6,3.76923)(5.94728,3.61107)(6,3.61107)
\color{rgb_0000ff}%
\path(4.7917,0.461538)(4.80378,0.469994)(4.81586,0.478458)
  (4.82794,0.486939)(4.84003,0.495445)(4.85211,0.503984)
  (4.86419,0.512565)(4.87628,0.521197)(4.88836,0.529887)
  (4.90044,0.538645)(4.91253,0.547479)(4.92461,0.556398)
  (4.93669,0.56541)(4.94877,0.574525)(4.96086,0.583752)
  (4.97294,0.593099)(4.98502,0.602576)(4.99711,0.612192)
  (5.00919,0.621957)(5.02127,0.63188)(5.03336,0.641971)
  (5.04544,0.652241)(5.05752,0.662699)(5.06961,0.673355)
  (5.08169,0.68422)(5.09377,0.695305)(5.10585,0.706621)
  (5.11794,0.718179)(5.13002,0.72999)(5.1421,0.742066)
  (5.15419,0.754419)(5.16627,0.767061)(5.17835,0.780005)
  (5.19044,0.793262)(5.20252,0.806848)(5.2146,0.820774)
  (5.22668,0.835055)(5.23877,0.849704)(5.25085,0.864737)
  (5.26293,0.880167)(5.27502,0.896011)(5.2871,0.912284)
  (5.29918,0.929001)(5.31127,0.94618)(5.32335,0.963837)
  (5.33543,0.98199)(5.34752,1.00066)(5.3596,1.01986)(5.37168,1.03961)
  (5.38376,1.05993)(5.39585,1.08084)(5.40793,1.10236)
  (5.42001,1.12452)(5.4321,1.14732)(5.44418,1.17081)(5.45626,1.195)
  (5.46835,1.21991)(5.48043,1.24557)(5.49251,1.272)(5.5046,1.29923)
  (5.51668,1.32729)(5.52876,1.35621)(5.54084,1.386)(5.55293,1.41671)
  (5.56501,1.44836)(5.57709,1.48099)(5.58918,1.51462)(5.60126,1.5493)
  (5.61334,1.58504)(5.62543,1.6219)(5.63751,1.6599)(5.64959,1.69908)
  (5.66167,1.73948)(5.67376,1.78115)(5.68584,1.82412)
  (5.69792,1.86843)(5.71001,1.91413)(5.72209,1.96127)
  (5.73417,2.00988)(5.74626,2.06003)(5.75834,2.11175)(5.77042,2.1651)
  (5.78251,2.22013)(5.79459,2.2769)(5.80667,2.33546)(5.81875,2.39587)
  (5.83084,2.45819)(5.84292,2.52248)(5.855,2.5888)(5.86709,2.65723)
  (5.87917,2.72782)(5.89125,2.80065)(5.90334,2.87578)(5.91542,2.9533)
  (5.9275,3.03328)(5.93958,3.1158)(5.95167,3.20093)(5.96375,3.28878)
  (5.97583,3.37941)(5.98792,3.47292)(6,3.5694)(6.01208,3.47292)
  (6.02417,3.37941)(6.03625,3.28878)(6.04833,3.20093)(6.06042,3.1158)
  (6.0725,3.03328)(6.08458,2.9533)(6.09666,2.87578)(6.10875,2.80065)
  (6.12083,2.72782)(6.13291,2.65723)(6.145,2.5888)(6.15708,2.52248)
  (6.16916,2.45819)(6.18125,2.39587)(6.19333,2.33546)(6.20541,2.2769)
  (6.21749,2.22013)(6.22958,2.1651)(6.24166,2.11175)(6.25374,2.06003)
  (6.26583,2.00988)(6.27791,1.96127)(6.28999,1.91413)
  (6.30208,1.86843)(6.31416,1.82412)(6.32624,1.78115)
  (6.33833,1.73948)(6.35041,1.69908)(6.36249,1.6599)(6.37457,1.6219)
  (6.38666,1.58504)(6.39874,1.5493)(6.41082,1.51462)(6.42291,1.48099)
  (6.43499,1.44836)(6.44707,1.41671)(6.45916,1.386)(6.47124,1.35621)
  (6.48332,1.32729)(6.4954,1.29923)(6.50749,1.272)(6.51957,1.24557)
  (6.53165,1.21991)(6.54374,1.195)(6.55582,1.17081)(6.5679,1.14732)
  (6.57999,1.12452)(6.59207,1.10236)(6.60415,1.08084)
  (6.61624,1.05993)(6.62832,1.03961)(6.6404,1.01986)(6.65248,1.00066)
  (6.66457,0.98199)(6.67665,0.963837)(6.68873,0.94618)
  (6.70082,0.929001)(6.7129,0.912284)(6.72498,0.896011)
  (6.73707,0.880167)(6.74915,0.864737)(6.76123,0.849704)
  (6.77332,0.835055)(6.7854,0.820774)(6.79748,0.806848)
  (6.80956,0.793262)(6.82165,0.780005)(6.83373,0.767061)
  (6.84581,0.754419)(6.8579,0.742066)(6.86998,0.72999)
  (6.88206,0.718179)(6.89415,0.706621)(6.90623,0.695305)
  (6.91831,0.68422)(6.93039,0.673355)(6.94248,0.662699)
  (6.95456,0.652241)(6.96664,0.641971)(6.97873,0.63188)
  (6.99081,0.621957)(7.00289,0.612192)(7.01498,0.602576)
  (7.02706,0.593099)(7.03914,0.583752)(7.05123,0.574525)
  (7.06331,0.56541)(7.07539,0.556398)(7.08747,0.547479)
  (7.09956,0.538645)(7.11164,0.529887)(7.12372,0.521197)
  (7.13581,0.512565)(7.14789,0.503984)(7.15997,0.495445)
  (7.17206,0.486939)(7.18414,0.478458)(7.19622,0.469994)(7.2083,0.461538)
\put(5.96485,0.426392){\makebox(0,0)[tr]{\hbox{\color{rgb_000000}\scriptsize $0$}}}
\put(6,3.83952){\makebox(0,0)[b]{\hbox{\color{rgb_000000}\scriptsize $y$}}}
\put(7.7626,0.461538){\makebox(0,0)[l]{\hbox{\color{rgb_000000}\scriptsize $\theta$}}}
\put(4.7917,0.391247){\makebox(0,0)[t]{\hbox{\color{rgb_000000}\scriptsize $-\frac{\pi}{4}$}}}
\put(7.2083,0.391247){\makebox(0,0)[t]{\hbox{\color{rgb_000000}\scriptsize $\frac{\pi}{4}$}}}
\put(6.37798,1.84766){\makebox(0,0)[l]{\hbox{\color{rgb_000000}\scriptsize \textcolor{blue}{$y=\Theta_{\lambda_{0}}(\theta)$}}}}
\end{picture}%